\documentclass[twocolumn]{aastex701}
\usepackage{hyperref}
\received{2026 May 15}
\revised{2026 October 9}
\accepted{2026 October 12}

\submitjournal{ApJ}

\shorttitle{RR Lyrae in Leo~P}
\shortauthors{Slaughter et al.}

\usepackage{amsmath}
\usepackage{float}

\begin{document}

\title{JWST Observations of RR Lyrae in Leo~P}

\author[0000-0002-5752-3780]{Catherine M.\ Slaughter}
\email{slaug098@umn.edu}
\altaffiliation{Vising Pre-Doctoral Fellow at the Center for Interdisciplinary Exploration and Research in Astrophysics, Northwestern University, 1800 Sherman Avenue, Evanston, IL, 60201, USA}
\affiliation{Minnesota Institute for Astrophysics, University of Minnesota, Minneapolis, MN 55455, USA}

\author[0000-0002-1445-4877]{Alessandro Savino}
\email{astrosavino@gmail.com}
\affiliation{Department of Astronomy, University of California, Berkeley, Berkeley, CA 94720, USA}

\author[0000-0001-7531-9815]{Meredith Durbin}
\email{mdurbin@uw.edu}
\affiliation{Department of Astronomy, University of Washington, Box 351580, Seattle, WA 98195-1580, USA}
\affiliation{Department of Astronomy, University of California, Berkeley, Berkeley, CA 94720, USA}

\author[0000-0001-8416-4093]{Andrew E.\ Dolphin}
\email{adolphin@rtx.com}
\affiliation{Raytheon, 1151 E. Hermans Rd.,Tucson, AZ 85756}
\affiliation{Steward Observatory, University of Arizona, 933 N. Cherry Avenue, Tucson, AZ 85719, USA}

\author[0000-0001-5538-2614]{Kristen B.\ W.\ McQuinn}
\email{kmcquinn@stsci.edu}
\affiliation{Department of Physics and Astronomy, Rutgers, the State University of New Jersey,  136 Frelinghuysen Road, Piscataway, NJ 08854, USA}
\affiliation{Space Telescope Science Institute, 3700 San Martin Drive, Baltimore, MD 21218, USA}

\author[0000-0002-8092-2077]{Max J.\ B.\ Newman}
\email{mjbnewman25astro@gmail.com}
\affiliation{Department of Physics and Astronomy, Rutgers, the State University of New Jersey,  136 Frelinghuysen Road, Piscataway, NJ 08854, USA}
\affiliation{Space Telescope Science Institute, 3700 San Martin Drive, Baltimore, MD 21218, USA}

\author[0000-0003-0605-8732]{Evan D.\ Skillman}
\email{skill001@umn.edu}
\affiliation{Minnesota Institute for Astrophysics, University of Minnesota, Minneapolis, MN 55455, USA}

\author[0000-0002-6442-6030]{Daniel R.\ Weisz}
\email{dan.weisz@berkeley.edu}
\affiliation{Department of Astronomy, University of California, Berkeley, Berkeley, CA 94720, USA}

\begin{abstract}
We analyze co-spatial HST/ACS and JWST/NIRCam time-series photometric observations of RR Lyrae in the low-metallicity dwarf galaxy Leo~P. This study provides a new opportunity to explore the properties of and potential for metal-poor RR Lyrae studies with JWST using well-understood relationships from past HST work. We identify 20 RR Lyrae in the HST imaging and 15 RR Lyrae in the JWST imaging. We determine the periods, amplitudes, and mean magnitudes for each of the RR Lyrae.  We fit the Period-Wesenheit-Metallicity (PWZ) function to the HST RR Lyrae population, and obtain a \textit{Gaia}-consistent RR Lyrae-based distance to Leo~P of $\mu = 26.04\pm0.07$ ($1.61\pm0.06$ Mpc), consistent with existing distance measurements. We determine a JWST-based PWZ using Leo~P data along with previous JWST RR Lyrae observations in WLM. The simultaneous fit decreases the mean fractional uncertainty by 31\% on the slope and by 11\% on the intercept, increasing the constraining power as compared to the existing calibration.
\end{abstract}


\section{Introduction} \label{sec:intro}

\subsection{RR Lyrae in Resolved Populations Studies}

RR Lyrae are a class of horizontal branch (HB) stars which fall within the pulsational instability strip of the HR diagram. As such, a resolved population of these sources can serve as a standard-candle distance indicator, making use of the correlation between their pulsational periods and intrinsic brightnesses. Modern high-resolution telescopes continue to expand the volume of possible studies of resolved stellar populations well beyond the reach of the Gaia mission \citep[e.g.,][]{Gaia2016, BailerJones2021}, and to increasingly distant extragalactic systems. RR Lyrae have been prolific in studies of Galactic archaeology \citep[e.g.,][]{Brown2003, Monelli2022, DOrazi2024, Bono2026}, Andromeda and its satellites \citep[e.g.,][]{Brown2004, Monelli2017, MartinezVazquez2017, Savino2022}, and other local dwarf galaxies \citep[e.g.,][]{Bernard2008, Bernard2009, Bernard2010, Bernard2013, McQuinn2015}, including clusters in the Magellanic Clouds \citep[e.g.,][]{Walker1989}.

In addition to serving as standard candles, RR Lyrae can provide an independent constraint on the oldest star formation rates in their host galaxies. As HB stars, RR Lyrae are, with few exceptions \citep{Bobrick2024}, at least $10$ Gyr old \citep{Walker1989, Savino2020, Monelli2022}. Thus, the existence of RR Lyrae in a stellar population is indicative of ancient star formation. Robust RR Lyrae identification may serve as a useful tool for studying star formation rates in resolved populations, for example, those in dwarf galaxies.



\subsection{Studying RR Lyrae in JWST Archival Data}

JWST is an immensely powerful tool for resolved stellar population studies in the local volume. Many such studies require relatively long integration times, from which time-series data can be extracted for potential use in variable star research. Of course, archival research such as this presents some expected challenges related to a lack of control over observation schema, as well as the JWST-specific filter set.

\citet{Slaughter2025} outlines the primary expected difficulties with studying RR Lyrae using JWST. They discuss how the amplitudes of the variations in RR Lyrae stars are reduced in IR data. Additionally, they highlight the impact of limitations on filter switching while continuously observing a target, in order to better preserve the NIRCam filter wheel.

Using coincident observations of the dwarf irregular galaxy WLM in HST and JWST, \citet{Slaughter2025} attempt to address these potential concerns. They outline the process for obtaining the necessary time sampling from the JWST archive. They determine that while variability identification remains robust, existing standard period-measurement and RR Lyrae classification methodology falls short for relatively long-period RR Lyrae sources in short-duration data. By anchoring their JWST analysis to HST-derived period values, they determine that the reduced amplitude and continuous-filter effects should not negatively impact the template-fitting procedure, given adequate phase sampling. Finally, they present a first-of-its-kind calibration of the fundamental-mode RR Lyrae (RRab) Period-Wesenheit-Metallicity function in JWST bands.

Here, we use similar existing coincident HST and JWST observations of the nearby low-metallicity dwarf galaxy Leo~P to characterize JWST's RR Lyrae detection and classification efficiency on a longer-baseline dataset. We use these Leo~P data to supplement the WLM dataset with more sources, in order to better constrain the existing PWZ calibration in the JWST bands.

\subsection{The Low-Mass Star-Forming Galaxy Leo~P}

Leo~P is a gas-rich, star-forming dwarf galaxy. Its relative isolation makes Leo~P a important probe of low-mass galaxy formation. While we expect there to exist many such isolated low-mass systems, their detection continues to be difficult, largely due to their low surface brightness. A thorough study of the RR Lyrae in Leo~P, therefore, would benefit our understanding of RR Lyrae as probes of very old star formation in such systems.

Using HST data, \cite{McQuinn2015} identify a small population of 10 RR Lyrae candidates in Leo~P. They report on-sky locations; mean magnitudes in the F475W, F814W, and (transformed) Johnson V bands; and periods for each source. Here, we make use of the same HST data, allowing for a direct comparison of variability-identification and period-determination results.

In the same study, \cite{McQuinn2015} also employ a number of standard-candle methods to measure the distance to Leo~P, including tip of the red giant branch ($\mu=26.19_{-0.50}^{+0.17}$), HB ($\mu=26.05\pm0.20$), and RR Lyrae magnitude-metallicity relationships (using the Johnson V band and [Fe/H] metallicity; $\mu=26.04 \pm 0. 21$). They ultimately adopt the HB measurement, corresponding to a physical distance of $1.62 \pm 0. 15$ Mpc. 

\citet{McQuinn2024LeoP} make use of deep JWST photometry (also utilized here) in order to obtain a robust star formation history (SFH) and age-metallicity relation (AMR) for Leo~P based on a full sampling of the oldest main sequence turnoff (oMSTO). The AMR results show that there has been very little chemical evolution in Leo P, with the average metal abundance ranging between $\langle\text{[M/H]}\rangle = -1.4$ to $-1.2$. The SFH they derive shows an early period of star formation prior to the Epoch of Reionization, a pause following this epoch (from z$\sim$5 to 1), and later re-ignition of new star formation. Leo~P's SFH is therefore comparable to other dwarf galaxies which exist in a ``transition zone'' between low-mass, quenched galaxies, and high-mass galaxies with SFHs that appear to have been unaffected by reionization. They highlight that existing cosmological simulations can contain these transition-zone galaxies, but they are uncommon.


\section{Data}\label{sec:data}

This paper makes use of archival data from both the HST and JWST archives, first presented in \citet{McQuinn2015} and \citet{McQuinn2024LeoP}, respectively.

\subsection{HST Observations}
HST Advanced Camera for Surveys \citep[ACS,][]{Ford1998} observations of Leo~P were taken between April 23-26, 2014 (GO 13376; PI McQuinn). The program was designed to achieve a depth $\sim2$ mag below the red clump in the F475W and F814W filters. The observations were taken over 17 orbits, grouped in visits of 2 to 3 orbits, to allow for adequate coverage of short-period variable stars. Each orbit was split into one F475W and one F814W image. The filters were observed in reverse order during the following orbit(s) for improved cadence coverage. 

For our purposes, we make use of the pre-processed photometric catalogs produced by \citet{McQuinn2015}. The combined catalog of variable and non-variable sources from these, plotted on a CMD, can be seen in \autoref{fig:fullcmds}.

\subsection{JWST Observations}
JWST NIRCam \citep[][]{Rieke2023} observations of Leo~P were taken in two visits between April 8-12, 2023 (GO-1617; PI McQuinn). The program was designed to achieve a depth below the oMSTO in the F090W (18.37 hr) and F150W (11.02 hr) filters. The observations utilize the same \texttt{ORIENT} setting between visits, as well as the standard 7-dither subpixel dither pattern, in order to ensure precise point-spread function (PSF) photometry, increase signal-to-noise (S/N), and improve sampling for the short-period variable stars. 

In order to obtain the necessary time sampling for RR Lyrae from the JWST imaging, we conduct the same JWST pre-processing as outlined in \citet{Slaughter2025}. In short, we are able to  extract the individual integration data of the Stage 1 data products (obtained from MAST) using a local run of the Stage 2 imaging pipeline\footnote{\url{https://jwst-pipeline.readthedocs.io/en/latest/jwst/pipeline/calwebb\_image2.html}} as if the data were dedicated time-series observations (TSOs). From this, we split the 3D calibrated data cubes into a series of ``pseudo-cal" 2D images (and use the accompanying table of integration times to correct timestamps in the headers). The pseudo-cal images are suitable for use with the existing \texttt{DOLPHOT} PSF photometry routine.

\begin{figure}
    \centering
    \includegraphics[width=0.49\linewidth]{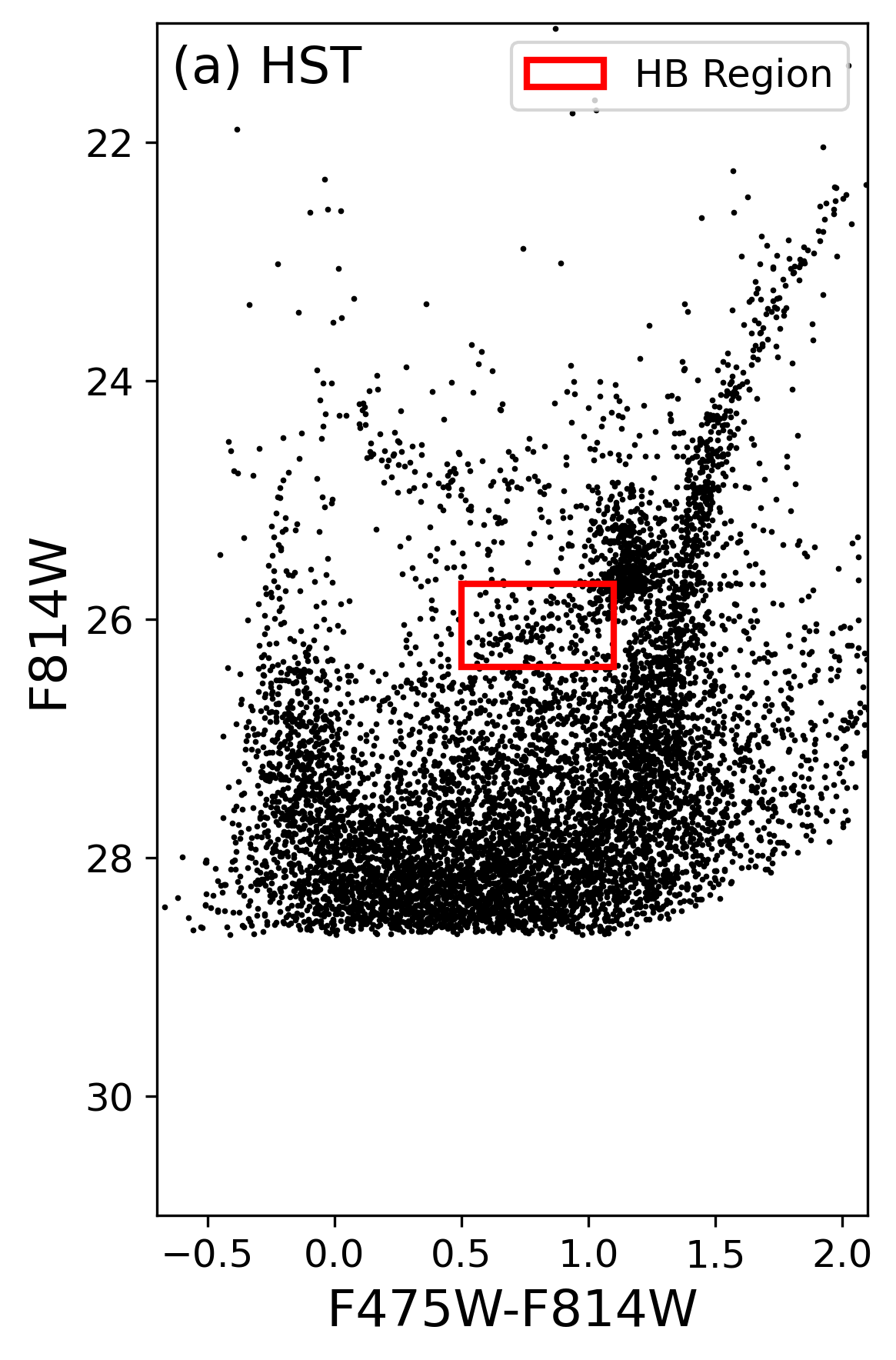}
    \includegraphics[width=0.49\linewidth]{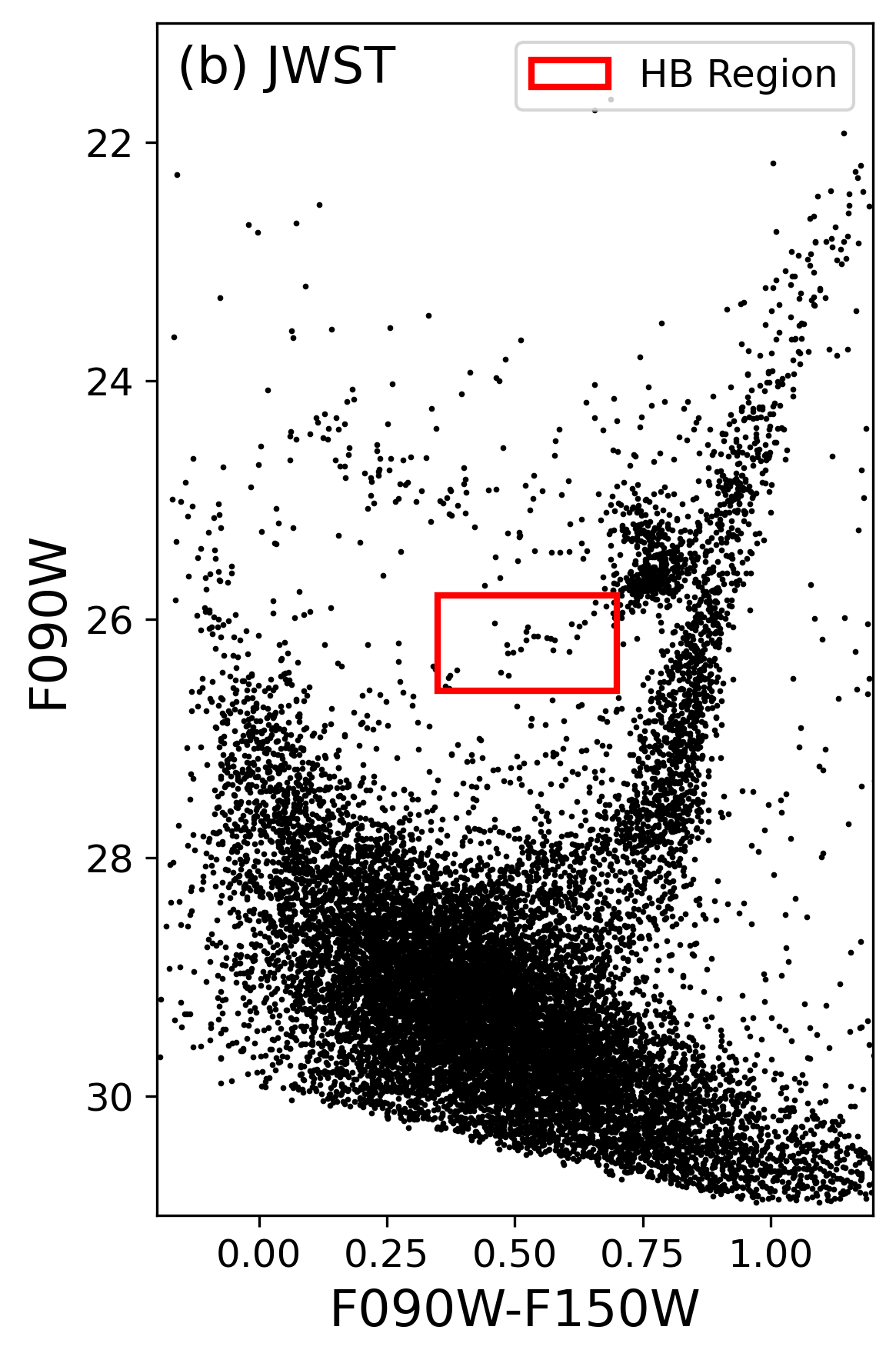}
    \caption{The quality-filtered photometric catalogs, on standard color-magnitude diagrams, for both the HST (left) and JWST (right) data. The horizontal branches (HB) can be seen in each just below and to the left of the red clump features. The red boxes outline the HBs, and  indicate the CMD regions used for RR Lyrae candidate selection in this study.}
    \label{fig:fullcmds}
\end{figure}

Hence, PSF photometry was performed using the \texttt{DOLPHOT} photometry package, which contains a NIRCam-specific module \citep{Weisz2024}. Filtered photometric catalogs of the point sources are created using the purity-oriented selection from \citet{Weisz2024}. In particular, we require the combined-band signal-to-noise ratio (SNR) $\ge 4$, the squared combined-band sharpness $\le 0.01$, and the combined-band crowding $\le 0.5$. The final catalog can be seen, plotted on the CMD, in \autoref{fig:fullcmds}.

\section{Methodology}\label{sec:methods}

The methodology we follow to detect and characterize RR Lyrae in the archival images follows \citet{Savino2022} and \citet{Slaughter2025}. Here we provide a brief summary of this methodology.

\subsection{Variable Source Identification}
\begin{figure}
    \centering
    \includegraphics[width=\linewidth]{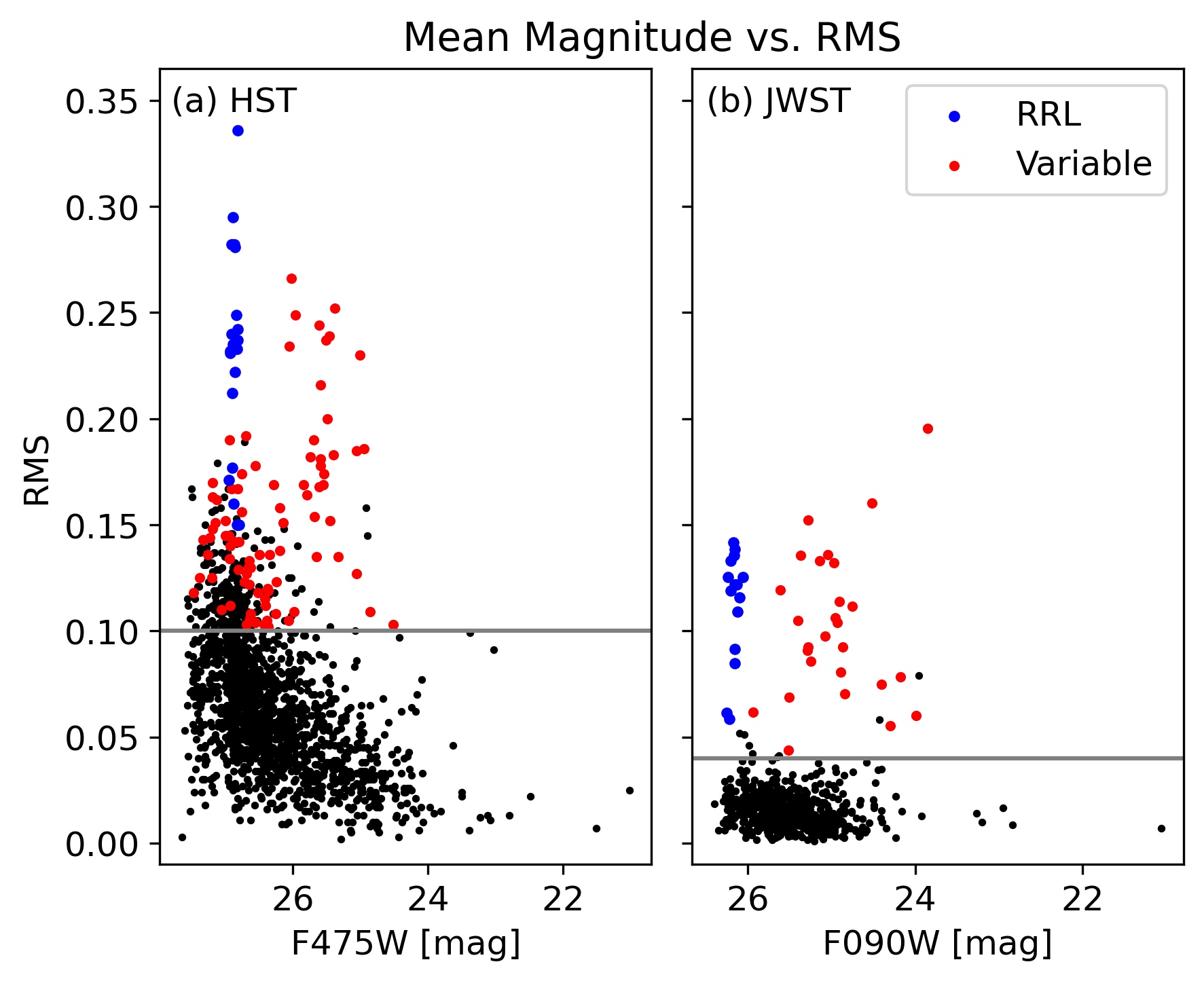}
    \caption{Plots showing the mean magnitudes vs. RMS variability for the HST (left) and JWST (right) observations. Note that only sources which pass the initial photometry and epoch number requirements are shown. The sources which are flagged as variable in the initial identification are shown in red, sources which are eventually confirmed to be RR Lyrae are shown in blue, while non-variable sources are shown in black. The RMS cutoffs (0.1, HST; 0.04, JWST) are indicated with gray horizontal lines. Sources which exceed the RMS threshold but are not flagged as variable are removed by either the $\chi^2$ or periodicity metric filters.}
    \label{fig:rmsmag}
\end{figure}

In order to identify RR Lyrae candidates from the photometry output, the catalog is first split into individual epochs. Light curves are obtained by pulling every individual epoch measurement for a given source. For each of these sources, an epoch-to-epoch comparison is conducted to assess the variability and periodicity in the light curve. We only consider sources detected in at least a given number of images (23 in HST; 75 in JWST), in order to assure sources with a small number of anomalous epochs do not contaminate the overall data set. The metrics to assess a given source's variability and periodicity are, in brief:
\begin{enumerate}
    \item The overall root-mean-squared (RMS) scatter must be greater than a given value. Based on the analysis from \citet{Slaughter2025} this value is set at 0.1 for the HST data, and 0.04 for the JWST data. This is to ensure the light curve exhibits some baseline variability. The choice of cutoff is illustrated in \autoref{fig:rmsmag}. In it, positive variability shows larger RMS than non-variable sources with similar magnitudes. Additionally, it shows that the selected RMS thresholds successfully cull the catalog without risking the removal of RR Lyrae sources.
    
    \item The $\chi^2$ metric of the variability must be greater than a given cutoff value for three cases: the full light curve, the light curve with the single most discrepant point removed, and the light curve with 1/3 of the most discrepant points removed. This is to determine if the observed variability is both statistically significant, and is not impacted by a handful of anomalous epochs. The $\chi^2$ metrics are combined between the two bands, as in \citet{McQuinn2015}. Following the analysis of \citet{Slaughter2025}, the cutoffs for the first two cases are 2.6 for the HST data and 1.0 for the JWST data. For both data sets, the third-pass cutoff is set to 0.5. 
    
    \item The minimum periodicity metric, $\Theta$, must be less than a given value of 0.8. This metric is calculated according to the Lafler-Kinman phase dispersion minimization algorithm \citep{Lafler1965} for test periods ranging from 0.25 to 3 days, i.e.: 
    \begin{equation}
        \Theta=\frac{\sum_{i=1}^N\left(m_i-m_{i+1}\right)^2}{\sum_{i=1}^N\left(m_i-\bar{m}\right)^2}.
    \end{equation}
    The individual metrics for the two filters are combined as in \cite{McQuinn2015}. This is to ensure that the observed variability also exhibits periodicity.
    
\end{enumerate}

As described in \citet{Slaughter2025}, the threshold values used are adjusted between the HST and JWST observations to account for the reduced amplitudes in the IR. Here, we adopt the same cutoffs they establish. As compared to the analysis done in \citet{McQuinn2015}, the variability thresholds (1 and 2 above) adopted here from \citet{Slaughter2025} are more relaxed. \citet{McQuinn2015} implements an RMS cutoff of 0.14 mag, and a first- and second-pass $\chi^2$ cutoff of 3.0. The third-pass $\chi^2$ cutoff and periodicity metric constraints are the same as above. This is meant to increase completeness in the candidate list, as contaminant sources are more effectively rejected at a later stage of the procedure.

Once a set of variable sources is identified, we inspect the CMD and isolate the candidates which lie within about 0.4 mag from the mean of the HB \citep[based on the work from][and visual inspection of the CMD]{Savino2022}, and the approximate color of the instability strip. \autoref{fig:fullcmds} shows the chosen HB regions over the catalog CMDs. For the HST bands, the isolated region is from 25.7 to 26.4 mag in F814W, and 0.4 to 1.1 in F475W-F814W color. For JWST, 25.8 to 26.6 mag in F090W, and 0.35 to 0.7 in F090W-F150W color. 

\subsection{Periodicity Measurement}
For each RR Lyrae candidate, we determine a first estimate of the period using the hybrid period-determination method from \citet{Saha2017}. This method, which has been optimized for sparsely-sampled, multi-band data, makes simultaneous use of the previously-mentioned Lafler-Kinman \citep{Lafler1965} algorithm, and a Lomb-Scargle algorithm \citep{Lomb1976, Scargle1982}. 

The Lafler-Kinman is a phase-dispersion minimization (PDM) algorithm, which works by comparing the relative values of neighboring data points in a phase-folded light curve for a variety of test periods. For a test period similar to the real period of the source, the magnitudes of adjacent data points in the associated phase space can be expected to be similar, giving the test period greater power in the final periodigram.

The Lomb-Scargle is a harmonic analysis (HA) algorithm, which works by fitting sine curves of various periods to the overall data. The power of a given period is determined by the amplitude of the best-fit sine curve with said period. This method is used in place of a full Fourier-Series analysis to avoid computationally-intensive data resampling.

Making use of both of these algorithms as a combined PDM-HA method has been shown to reduce the impact of artifacts introduced by the sampling in the observations \citep{Saha2017}.

\subsection{MCMC Template Fitting and Classification}

The final pulsation parameters (i.e., periods, amplitudes, and mean magnitudes) are obtained through Markov-Chain Monte Carlo (MCMC) template-fitting, following the approach outlined in \citet{Savino2022}.

We obtain Fundamental Mode (RRab) and First-Overtone (RRc) pulsator templates for the relevant photometric bands from the literature. For the HST data, we make use of Johnson B and I templates from \citep{Monson2017}, which map to the F475W and F814W bands, respectively. For the JWST data, we use similar templates for the standard I and H bands, which reasonably map to the F090W and F150W filters. The I-band templates again come from \citet{Monson2017}, while the H-band templates were created later by \citet{Braga2019}. For a given set of test pulsation parameters, model light curves are generated from the best-fit template for the associated filter.

\begin{figure}
    \centering
    \includegraphics[width=\linewidth]{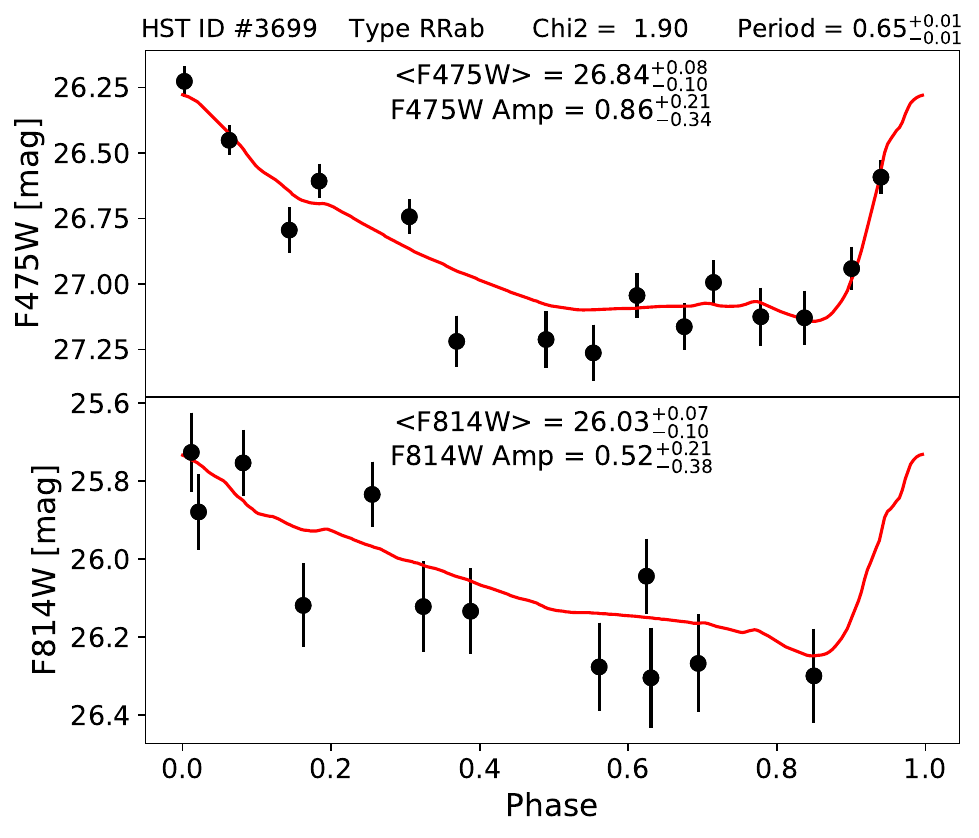}
    \caption{Example template fits for the HST source \#3699 for the F475W (top) and F814W (bottom) data.}
    \label{fig:tempfitHST}
\end{figure}

Using a Gaussian likelihood function and broad, uninformative priors, we use the affine-invariant ensemble MCMC sampler \texttt{emcee} \citep{Foreman-Mackey2013} to sample the posterior probability distributions for the set of model light curve fit parameters for each star. We define the convergence length of the MCMC chain as 100 times the autocorrelation length. From the MCMC posterior distribution, we obtain median periods, amplitudes, and mean magnitudes for each source with uncertainties, and identify the pulsation mode (based on the period and best-fitting template). Figure \autoref{fig:tempfitHST} shows an example template fit for an RR Lyrae source originally identified in \citet{McQuinn2015}. The full set of pulsation parameters and types can be found in \autoref{sec:datatable}, \autoref{tab:extendedtable}.

We then make use of the model fits and posterior distributions, as well as diagnostic plots such as period-amplitude (``Bailey'') diagrams and CMDs to visually inspect and address irregular sources. Any sources which exhibit obvious and consequential aliasing are rerun with a more constrained prior on the period. In addition, sources which have obviously been mis-typed in the template selection step (based on their periods and locations on the Bailey diagrams) are rerun with the alternate type enforced. Finally, we remove any remaining non-RR Lyrae, determined by visual inspection.

\subsection{PWZ Fitting}

Once we have the pulsation data for the RR Lyrae population, we conduct an MCMC fit of the HST-calibrated PWZ function to the HST-observed RRab sources in order to estimate a distance to Leo~P \citep[as in][]{Savino2022, Slaughter2025}. The general form of the PWZ is
\begin{equation}\label{eq:pwz_general}
    W = \mu + a + b\ \text{log(P)} + c\ \text{[Fe/H]}.
\end{equation}
 Where W are the source-wise Wesenheit magnitudes \citep{Madore1982}; $\mu$ is the distance modulus; P are the pulsation periods in days; [Fe/H] are the metallicities; and $a$, $b$, and $c$ are calibrated—empirically or theoretically—for a given set of two filters. For clarity, we will make use of subscripts to refer to the calibration constants for the HST and JWST filter sets used here.

The Wesenheit magnitudes are defined to be reddening-free:
\begin{equation}\label{eq:W_obs}
    W (X, X-Y) = X - R(X - Y)
\end{equation}
\begin{equation}
    R = A_X / E(X-Y)
\end{equation}
where $R$ is the total-to-selective dust absorption ratio, and X, Y are the mean magnitudes for a given star in the associated filters.
 
For HST-based distancing, \citet{Savino2022} includes an additional correction to the standard form, in order to ensure consistency between distances from theoretically and empirically-derived values of $a_{HST}$, $b_{HST}$, and $c_{HST}$:
 \begin{equation}\label{eq:deltamu}
    \Delta\mu = 0.23\, \text{log(P)} - 0.02\, \text{[Fe/H]} - 0.087\, (V-I).
\end{equation}
The inclusion of $\Delta\mu$ means that the determined distance is consistent with \textit{Gaia} RR Lyrae observations \citep{Nagarajan2022}.

For the JWST data, we can use the standard form of the PWZ (as in \autoref{eq:pwz_general}, without $\Delta\mu$), along with the HST-derived distance modulus, to obtain JWST-specific calibration values. Because the sources lack individual metallicity measurements, the $a_{JWST}$, and $c_{JWST}$ become degenerate, and are combined into a single term $\alpha = a_{JWST} + c_{JWST}\text{[Fe/H]}$, yielding a functional PWZ:
\begin{equation}\label{eq:pwzalphab}
    W_{JWST} = \mu_{HST} + \alpha + b_{JWST}\ \text{log(P)}.
\end{equation} 
Where $\mu_{HST}$ is given a subscript only to indicate that it is calculated from the HST data, we emphasize that distance remains filter-agnostic.

\citet{Slaughter2025} present a first attempt at this calibration, using data from WLM. Here, we strengthen this calibration by simultaneously fitting their existing WLM observations and the new Leo~P observations.

To conduct the PWZ fittings, we again use \texttt{emcee}, following the Gaussian Mixture Model (GMM) formalism first described in \citet{Savino2022}. Various sources—such as aliasing and mis-classification—may introduce contamination into our sample. The GMM formalism allows us to account for the effect of outlier stars without the use of hard sigma-clipping by modeling them as being members of a second ``false" contaminant population. The contaminant population is drawn from a normal distribution, $\mathcal{N}(\mu_f, \sigma_f)$, where both $\mu_f$ and $\sigma_f$ serve as free nuisance parameters in our model. Instead of enforcing a binary decision as to which population a given star belongs to, we employ a sigmoid-like probability function:
\begin{equation}\label{eq:prob}
    Q_k = \frac{1}{1+\text{exp}(-s(R_k-2))},
\end{equation}
where $R_k = |\mu_k-\bar{\mu}|/\sigma^\mu_k$ is the normalized absolute deviation from the average of the $\mu_k$ distribution. This significantly reduces the constraining power of outlier sources. With this, we can calculate the likelihood of a given parameter set:

\begin{equation}
    \begin{split}
        &p(\mu_k|\mu, \textrm{[Fe/H]}, \mu_f, \sigma_f, s) \\
        &= \frac{(1 - Q_k)}{\sigma^\mu_k (2\pi)^2}\text{exp}(-(\mu_k-\mu)^2 / 2(\sigma^\mu_k)^2 \\ 
        &+ \frac{Q_k}{\sigma_f (2\pi)^2}\text{exp}(-(\mu_k-\mu_f)^2 / 2(\sigma_f)^2
    \end{split}
\end{equation}

The priors used for the HST-distancing GMM can be found in \autoref{tab:hstpriors}. We highlight that the range of the {[}Fe/H{]} prior used here is consistent with the range of HST-derived metallicities given in \citet{McQuinn2015}. They find an AMR which yields metallicity values ranging from $\text{[M/H]} = -1.7$ to $-1.6$, and an RGB-derived iron abundance of $\text{[Fe/H]} = -1.8\pm0.1$. 

\begin{table*}
    \centering
    \caption{Prior Distributions for the HST Gaussian Mixture Model\label{tab:hstpriors}}
    
    \begin{tabular}{ccc}
        \hline\hline
        Param. & Prior & Description \\
        \hline
        $\mu$ & $\mathcal{U}(20,30)$ & Distance modulus of the galaxy \\
        {[}Fe/H{]} & $\mathcal{U}(-2.25,-1.25)$ & Metallicity of the RRL population \\
        $\mu_f$ & $\mathcal{U}(20,30)$ & Dist. modulus of the contaminants\\
        $\sigma_f$ & $\mathcal{U}(0,10)$ & Scatter of the contaminants\\
        s & $\mathcal{U}(1,10)$ & Steepness of the Q sigmoid \\
        \hline
    \end{tabular}
\end{table*}

As in \citet{Slaughter2025}, we adapt the GMM formalism to the JWST calibration by replacing $\mu$ with $\Delta W$, the difference between W as derived from photometric data (as in \autoref{eq:W_obs}), and as derived from the PWZ (as in \autoref{eq:pwzalphab}). This eliminates the need to implement a 2-dimensional GMM for $\alpha_{JWST}$ and $b_{JWST}$.

\section{HST Results and Distance Determination}\label{sec:hstresults}

\begin{figure}
    \centering
    \includegraphics[width=\linewidth]{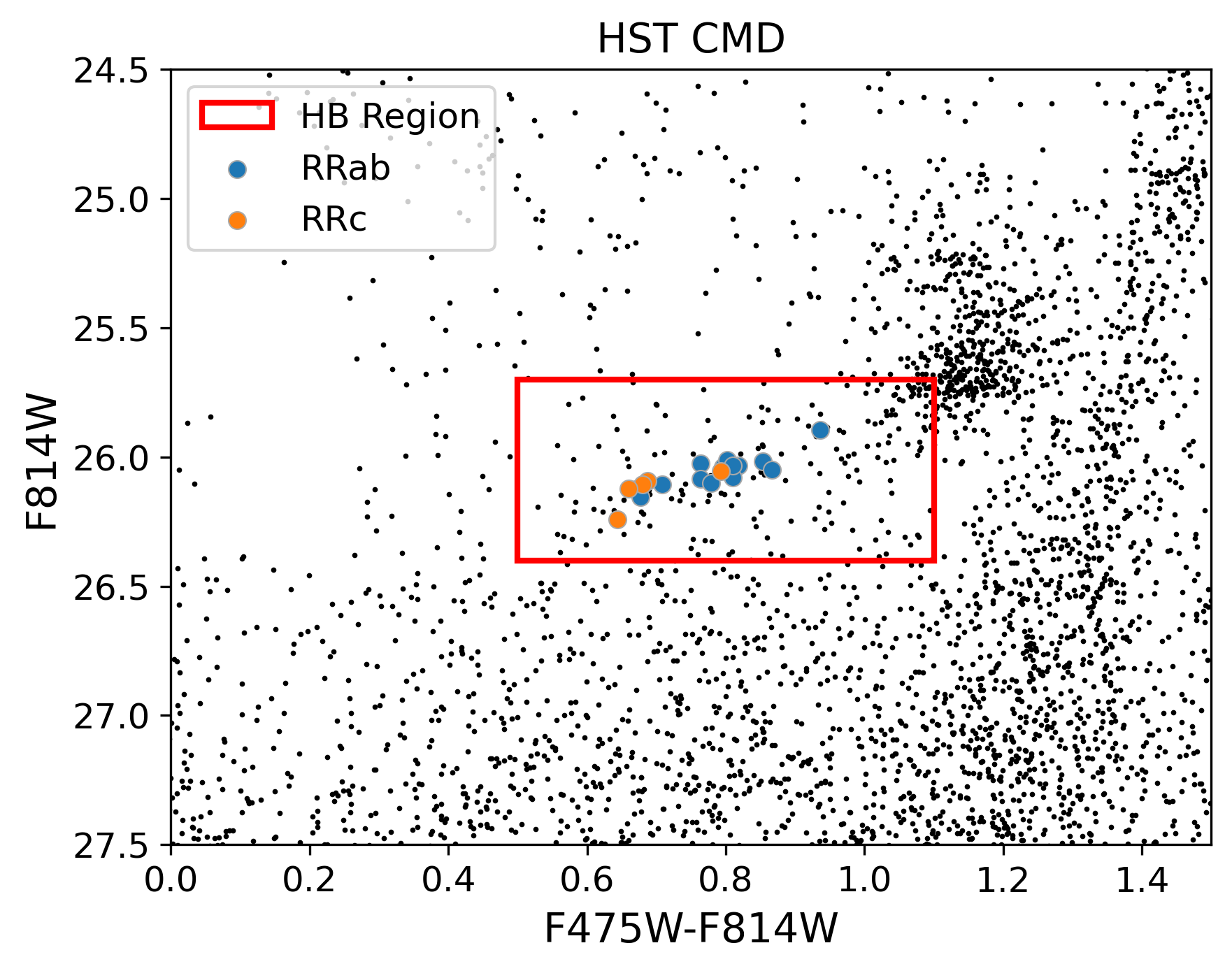}
    \caption{The CMD for the HST data, with the RR Lyrae sources indicated. RRab sources are in blue, and RRc sources are in orange. The region of the CMD used to search for RR Lyrae is outlined in red. In general, RRc sources have bluer mean colors than RRab, as expected. There is one RRc source that is notably redder than the rest, this may be an RRd source.}
    \label{fig:hstcmd}
\end{figure}

\begin{figure}
    \centering
    \includegraphics[width=\linewidth]{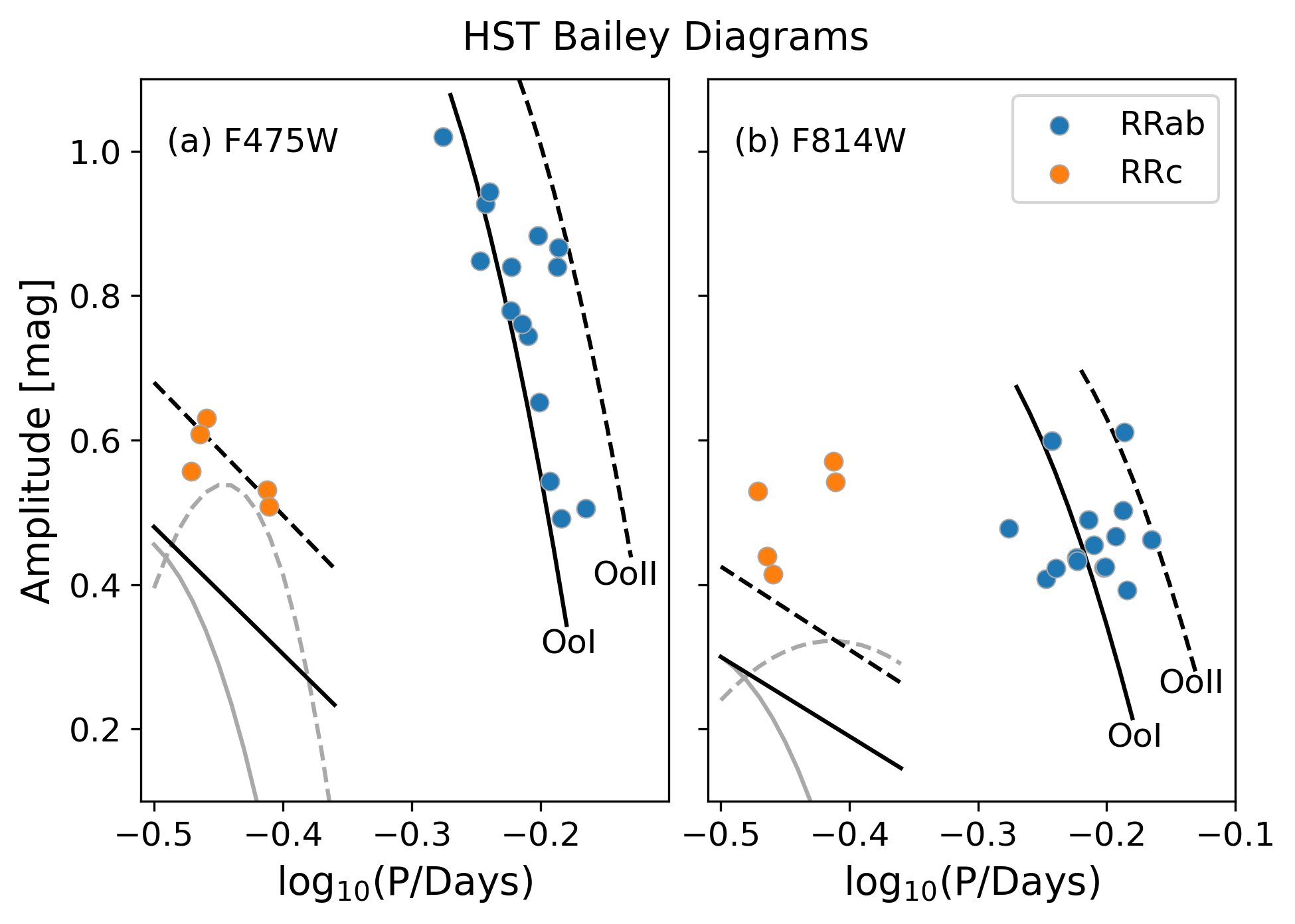}
    \caption{The Bailey diagrams for the HST F475W (left) and F814W (right) data. Note the separation between the RRab (blue) and RRc (orange) sources, where the RRab sources generally have greater amplitudes and longer periods. Lines show the empirical loci for Oosterhoff type I (solid) and Oosterhoff type II (dashed) globular clusters \citep[][]{Cacciari2005, ArellanoFerro2023}. V-band loci are plotted with the F475W data, and I-band loci with the F814W data. The black lines show loci for both RRab (parabolic) and RRc (linear) sources, reported in \citet{Kunder2013a} from observations of NGC 2808. For the RRc sources, more recent parabolic loci are also shown in gray. The Oosterhoff type I parabolic loci for both bands are reported in \citet{ArellanoFerro2015} using NGC 6229 observations. The V-band Oosterhoff type II parabolic locus is reported in \citet{Kunder2013b} using NGC 6656 observations. Finally, the I-band Oosterhoff type II parabolic locus is reported in \citet{Yepez2020} using NGC 6656 observations.}
    \label{fig:hstbailey}
\end{figure}

In the HST data, we identify 20 RR Lyrae, of which 15 are categorized as RRab sources. The final HST CMD and Bailey diagrams can be found in \autoref{fig:hstcmd} and \autoref{fig:hstbailey}, and show the familiar trends in periods, amplitudes, and colors between the RR Lyrae classes. There is some possibility that a small number of these sources are double-mode pulsators \citep[RRd,][]{Soszynski2016}. In sparsely-sampled data such as this, it is likely that these would be best fit as RRc sources, as RRd stars typically have primary periods associated with first-overtone (RRc) pulsation, and the amplitudes of the pulsations associated with the secondary periods are diminished compared to those of the first \citep[][]{Alcock2000, Soszynski2016, Soszynski2016b, Clementini2023, Nemec2024, Zhang2025}.

\begin{figure*}
    \centering
    \includegraphics[width=0.9\linewidth]{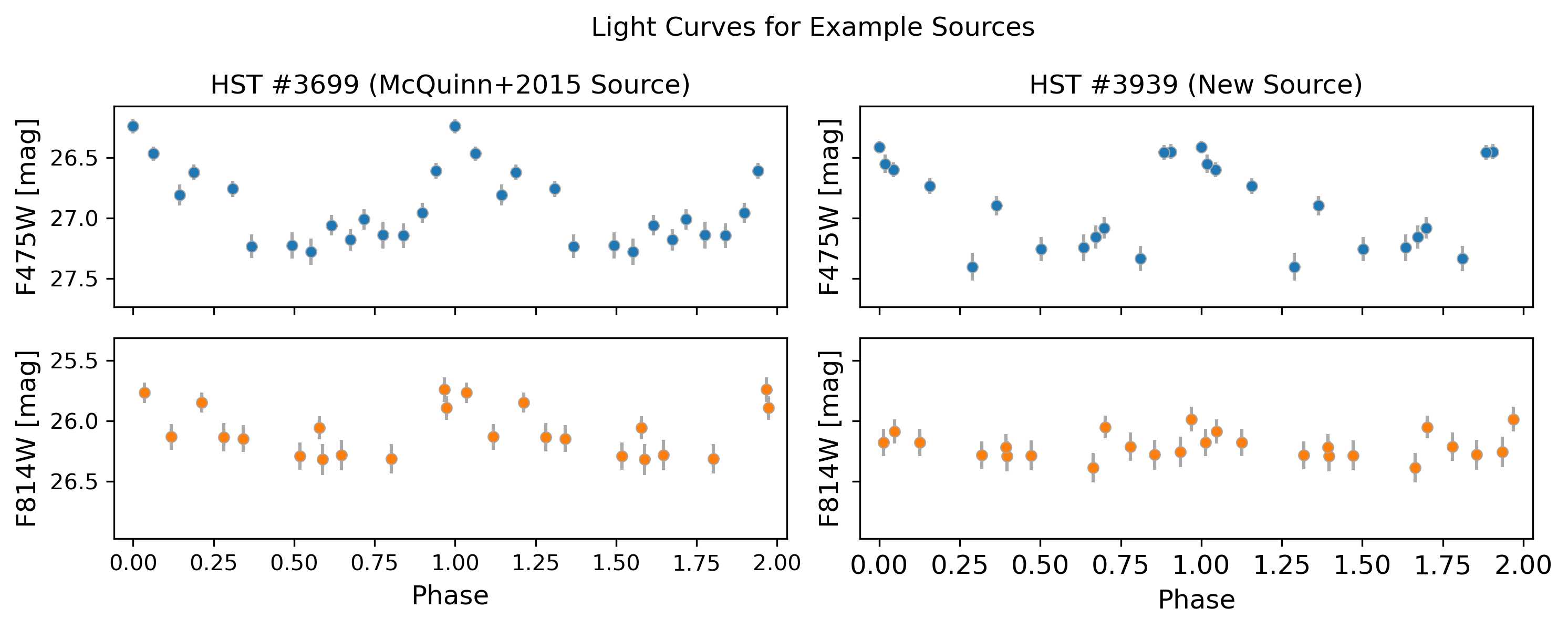}
    \caption{Light curve data (F475W, top; F814W, bottom) for two example RRab sources. Source \#3699 (left) is also identified as variable in \citet{McQuinn2015}, while source \#3939 (right) is new to this study. Source \#3699 is also used as an example in \autoref{fig:tempfitHST}}
    \label{fig:lightcurves}
\end{figure*}

We recover all 10 of the candidates identified in the results of \citet{McQuinn2015}, who searched for RR Lyrae in the same HST data. The additional sources in the catalog are found in large part due to relaxing the $\chi^2$ and RMS criteria in the variability identification step. In practice, the sources we recover that are not identified in \citet{McQuinn2015} tend to have larger photometric uncertainties relative to their amplitudes (this is especially clear in the relatively lower-amplitude F814W band), and a somewhat increased number of apparently discrepant epochs. Despite broad trends, we highlight that there does not appear to be one singular factor which differentiates the existing sources from the newly-identified ones, nor any clear cutoffs in the amplitude/period parameter space. Two sets of example RRab light curves \citep[one from a][source and one from a new source]{McQuinn2015} can be found in \autoref{fig:lightcurves}. The full suite of light curve figures can be found in \autoref{sec:fullLCs}, \autoref{fig:HSTlcFull}.

\begin{figure}
    \centering
    \includegraphics[width=\linewidth]{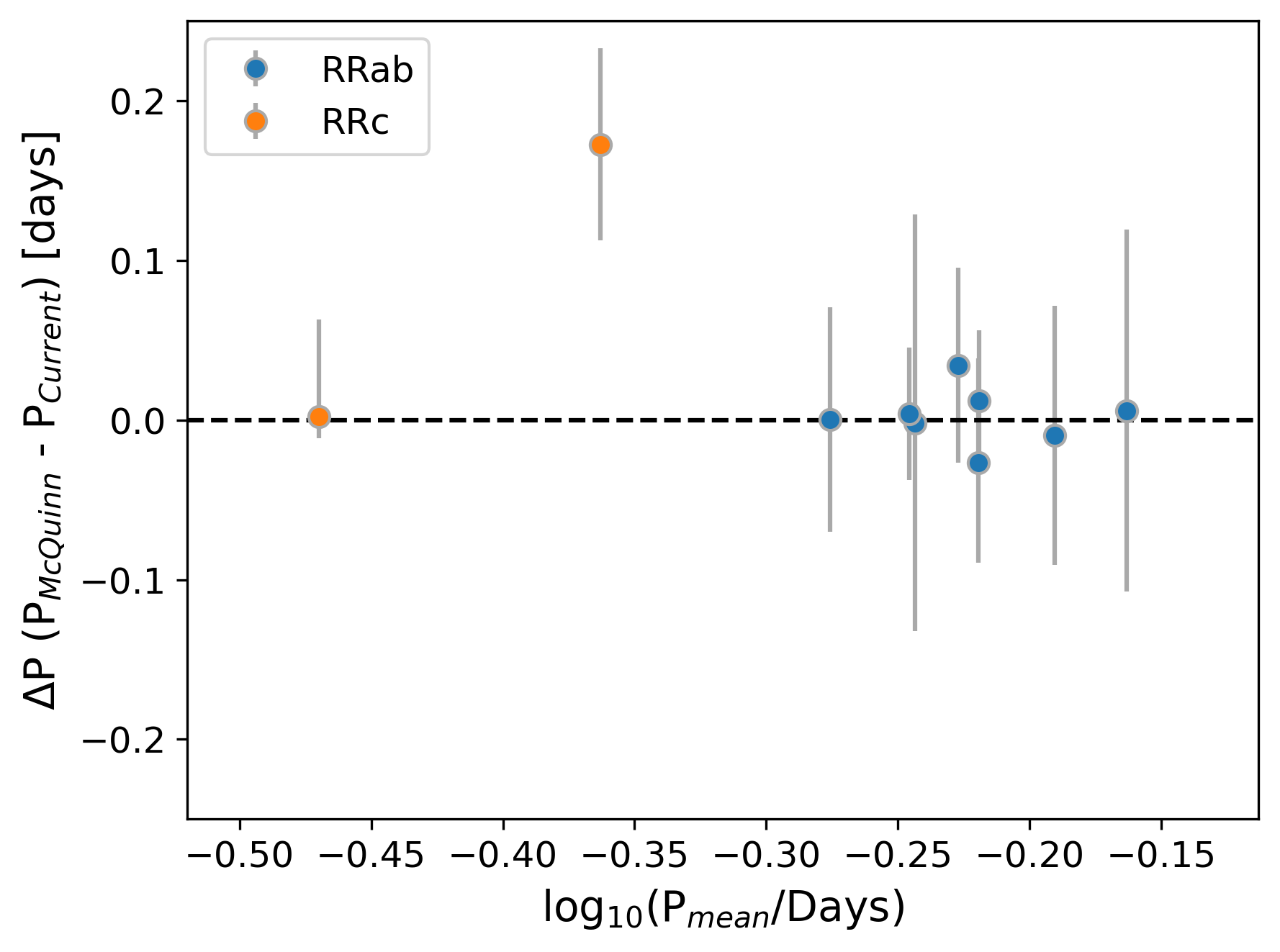}
    \caption{The difference in period estimates for the 10 RR Lyrae sources reported in \citet{McQuinn2015} and this work. Sources are colored by MCMC-derived classification (RRab, blue; RRc, orange). The black dashed line represents the expected difference of 0. Note that all sources are within 1-$\sigma$ error of the expected value.}
    \label{fig:PdiffsEV}
\end{figure}

We can compare the periods for the 10 \citet{McQuinn2015} sources to the MCMC template fitting. \autoref{fig:PdiffsEV} shows the average of the two period determinations versus the difference \citep[the][value minus the one reported in this work]{McQuinn2015} for all 10 sources, as well as their uncertainties (added in quadrature). We find strong agreement for 9 out of 10 sources, within 1-$\sigma$ uncertainty. The final source shows a 2.8-$\sigma$ difference. Upon examination, we find that the value reported in \citet{McQuinn2015} is more closely aligned with the first-guess period output by the Saha-Vivas algorithm here. Visual inspection of the MCMC fit indicates a few seemingly errant epochs in the F814W data which may skew the initial guess, a scenario which is notably well-handled by the MCMC. We also note no particular trend in $\Delta P$.

\begin{figure}
    \centering
    \includegraphics[width=\linewidth]{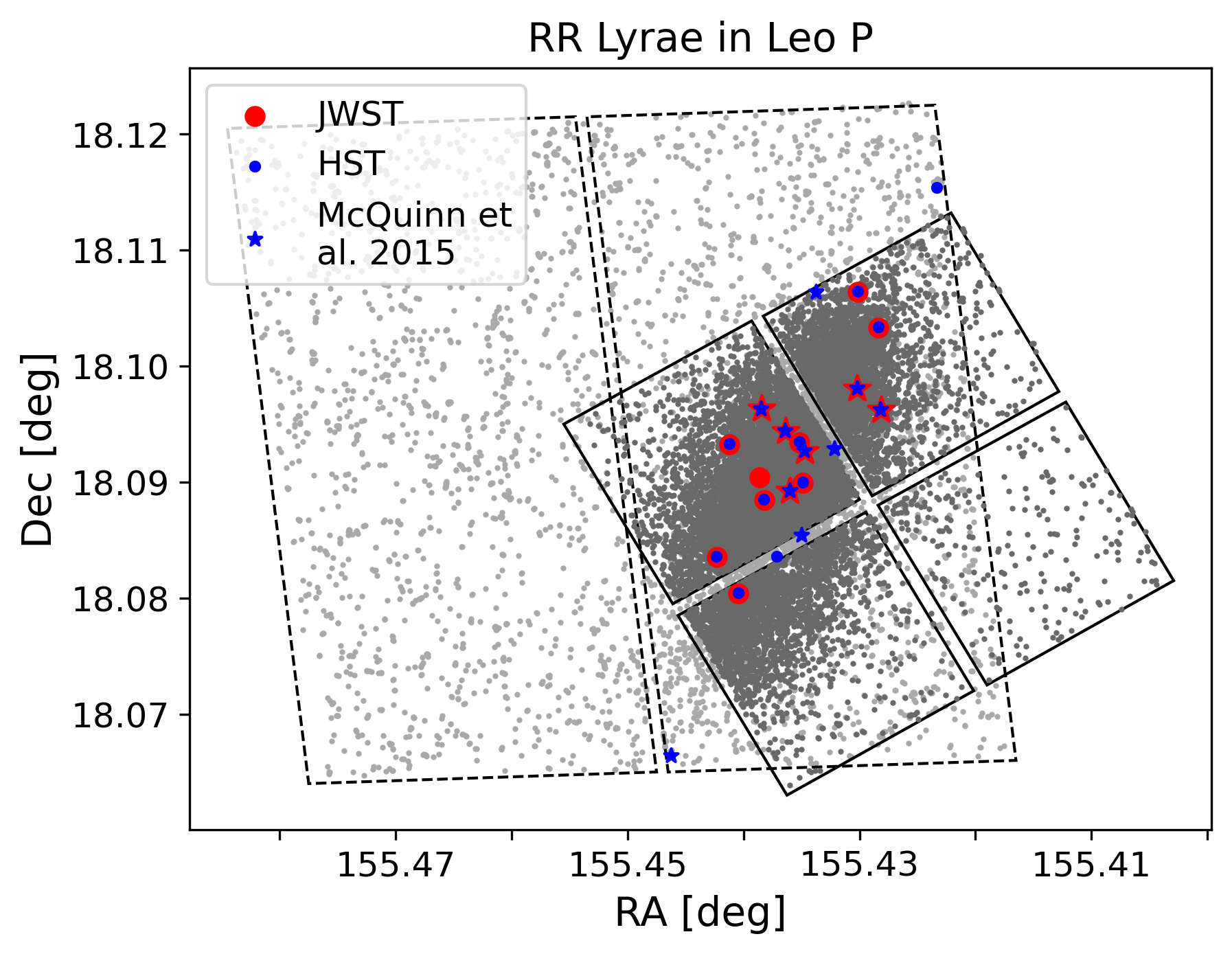}
    \caption{The RR Lyrae populations presented in this work from both HST (blue) and JWST (red), over-plotted with the rough outlines of the HST/ACS (dashed) and JWST/NIRcam (solid) footprints. The non-RR Lyrae sources from the HST (light gray) and JWST (dark gray) catalogs are also shown. Sources that were previously identified in \citet{McQuinn2015} are indicated as stars, while new sources are indicated as circles. Three of the McQuinn sources fall off the NIRCam detector. Note the one source in the coincident region which is seen by JWST but not HST.}
    \label{fig:overlap}
\end{figure}

\autoref{fig:overlap} shows the HST- and JWST-based RR Lyrae data sets, overlaid with the locations of the full stellar catalog (variable and non-variable sources; to demonstrate the overall density of sources) for each. In it, we can see the recovery of the \citet{McQuinn2015} sources in both HST and JWST (including where some fall off the NIRCam detector), as well as the recovery of the HST sources in JWST in the coincident region. There is 1 source in the coincident region which is found only in JWST and not HST. There is 1 additional source on the edge of the NIRCam chip that is found in HST, but not flagged as variable in JWST (see \autoref{fig:overlap}). This source is eliminated as a candidate early on in the variability-identification process because it does not show up in the PSF fitting for a sufficient number of epochs. This is most likely a combined effect of the dither pattern and poor PSF fit values from being on the edge of the image. The 5 other sources found in HST but not JWST are observed outside the JWST/NIRCam footprint.

\subsection{HST PWZ Distance Measurement}\label{sec:HSTpwz}

\begin{figure}
    \centering
    \includegraphics[width=\linewidth]{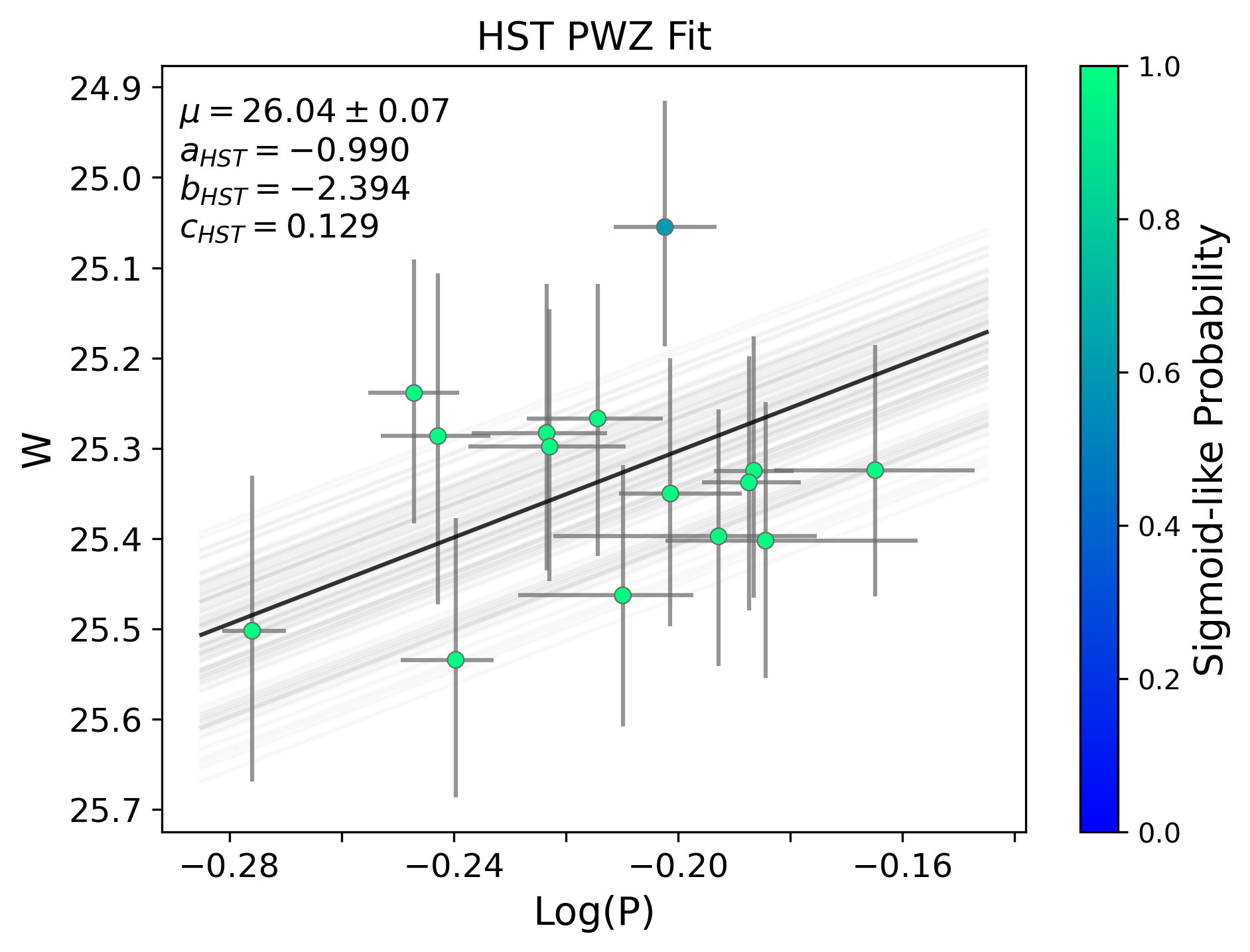}
    \caption{The PWZ fit to the HST data presented in this paper. The median fit line is presented in black, while a random subset of other MCMC fits are in gray. The sigmoid-like probability of belonging to the population generated by the median-fit line \citep[see][]{Savino2022,Slaughter2025} is indicated by color. The calculated distance modulus, as well as the PWZ calibration coefficients for the HST F475W and F814W bands are presented in the upper-left corner for reference. $a_{HST}$ is the standard offset, $b_{HST}$ is the period coefficient, and $c_{HST}$ is the metallicity coefficient.}
    \label{fig:hstpwz}
\end{figure}

Our HST PWZ fit is shown in \autoref{fig:hstpwz}. The median-fit line is shown in black, and a subset of other MCMC iterations are shown in gray, to highlight the spread. The data are colored according to the sigmoid-like probability of belonging to the RRab population. From this, we obtain a distance modulus to Leo~P of
$\mu = 26.04\pm0.07$ ($1.61\pm0.06$ Mpc).
This is in agreement with the \citet{McQuinn2015} RR Lyrae distance modulus \citep[$\mu_{\rm RRL} =26.04 \pm 0.21$; using the preferred calibration from][]{Carretta2000}, as well as their adopted HB distance modulus ($\mu_{\rm HB}=26.05 \pm 0.20$).

\section{JWST Results and Extended Distance Calibration}\label{sec:jwstresults}

\begin{figure}
    \centering
    \includegraphics[width=\linewidth]{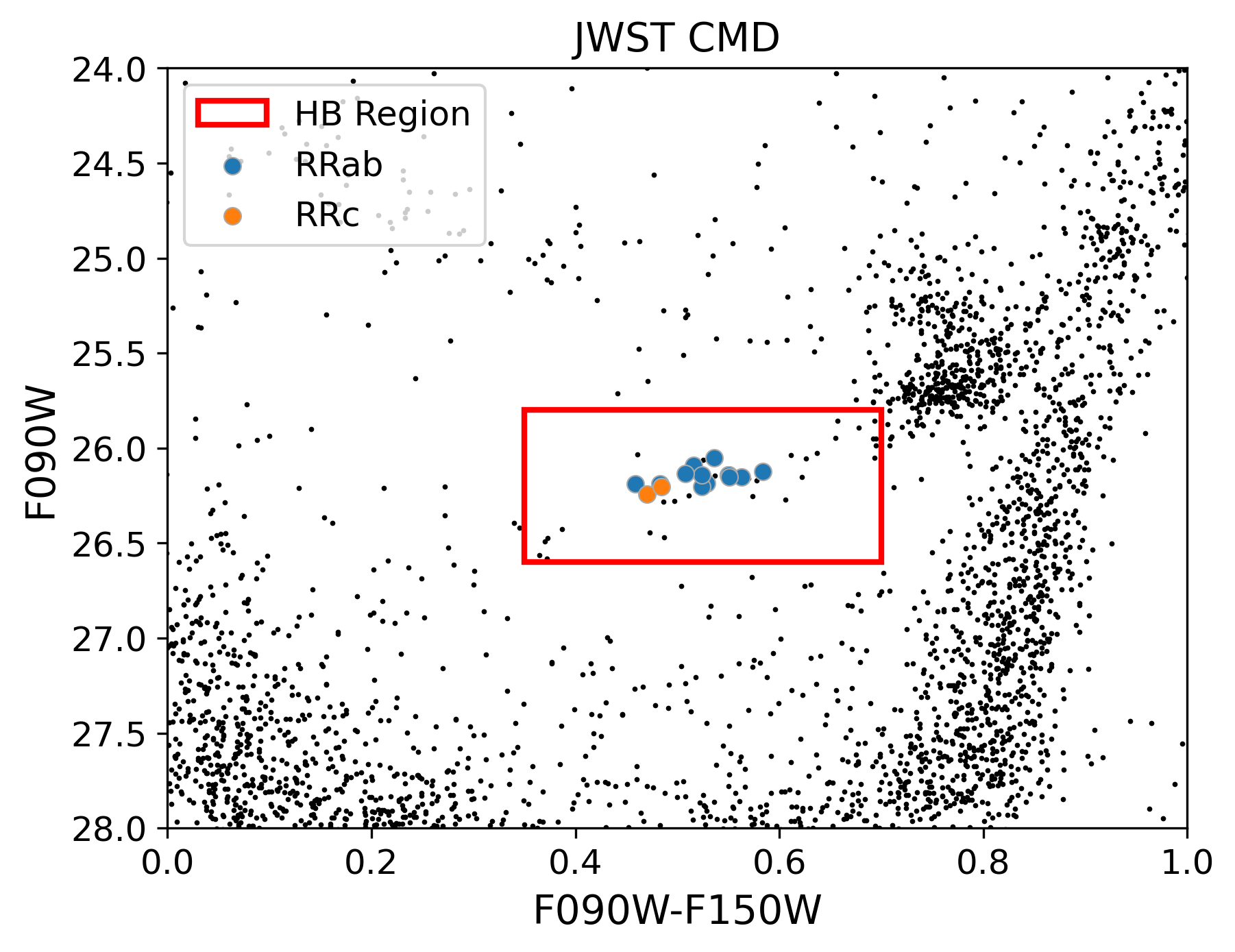}
    \caption{The CMD for the JWST data, centered around the HB. The RR Lyrae sources are highlighted. RRab in blue, RRc in orange. The general HB region is outlined in red.}
    \label{fig:jwstcmd}
\end{figure}

\begin{figure}
    \centering
    \includegraphics[width=\linewidth]{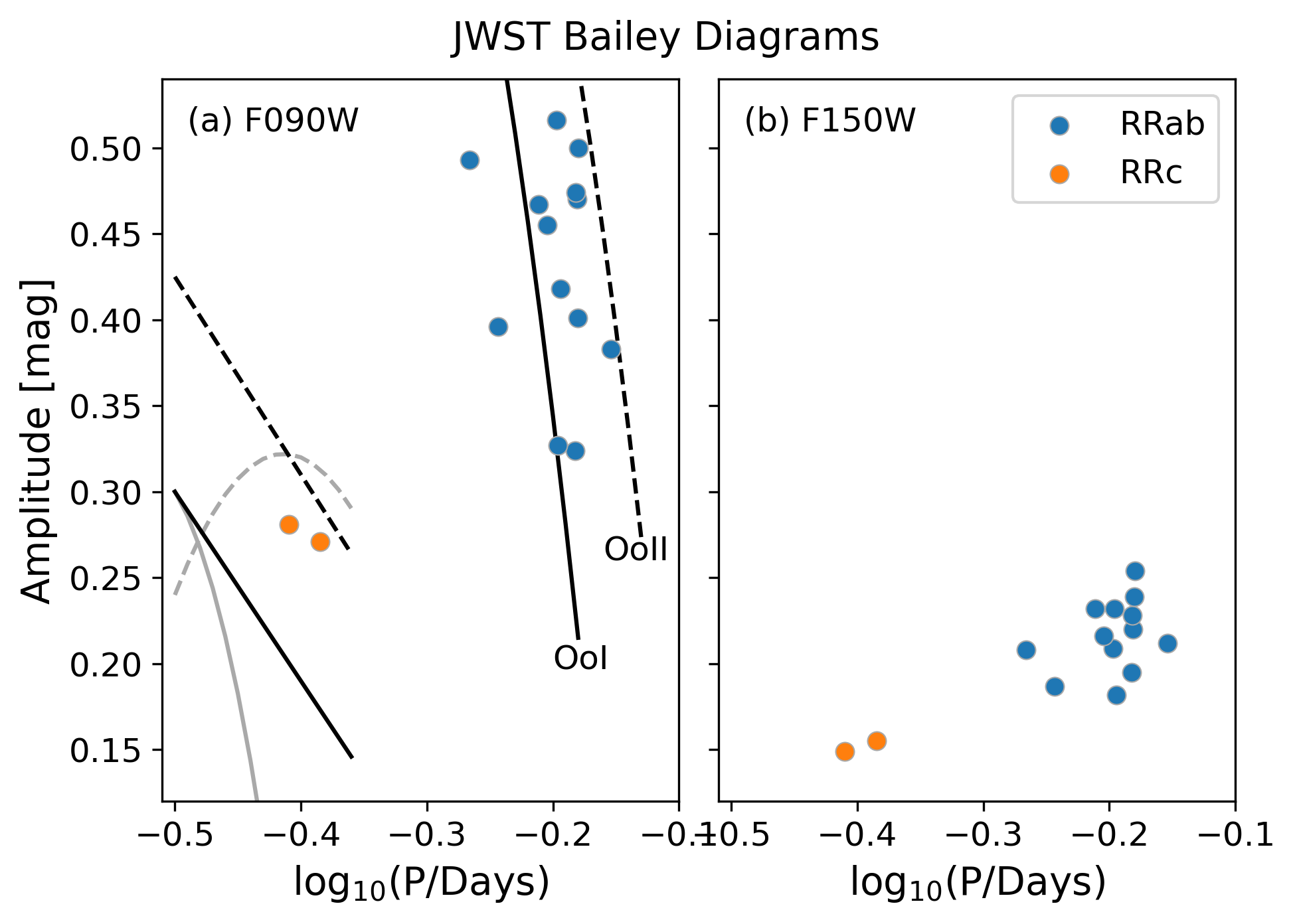}
    \caption{The Bailey diagrams made from the JWST  F090W (left) and F150W (right) data. Note the separation between the RRab (blue) and RRc (orange) populations. The RRcs have, on average, smaller amplitudes and shorter periods. Additionally note the reduced amplitudes in the F150W filter as compared to the bluer F090W filter. The I-band empirical loci for globular cluster Oosterhof classification from \autoref{fig:hstbailey} are plotted alongside the F090W data.}
    \label{fig:jwstbailey}
\end{figure}

In the JWST data, we identify 15 RR Lyrae, of which 14 of are co-identified in the HST data. The full suite of light curve figures can be found in \autoref{sec:fullLCs}, \autoref{fig:JWSTlcFull}. 13 of the 15 JWST RR Lyrae are categorized as RRab sources. The JWST CMD and Bailey Diagrams can be found in \autoref{fig:jwstcmd} and \autoref{fig:jwstbailey}. These again look generally as expected for a combined population of RRab and RRc sources.

\subsection{JWST PWZ Distance Measurement}

\begin{figure}
    \centering
    \includegraphics[width=\linewidth]{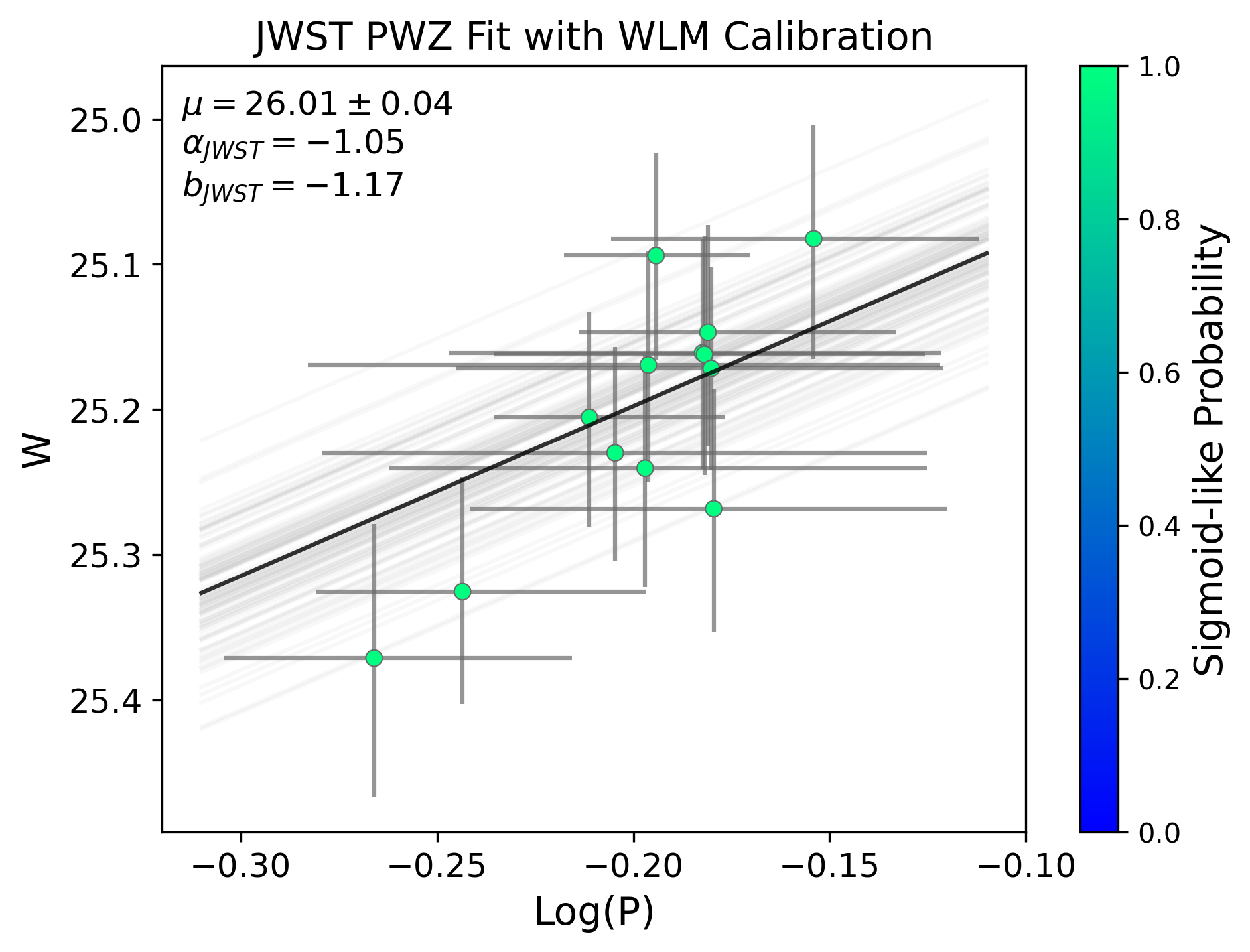}
    \caption{The PWZ fit to the JWST Leo-P data, using the WLM-derived PWZ calibration from \citet{Slaughter2025}. The median fit line is presented in black, while a random subset of other MCMC fits are in gray. The sigmoid-like probability of belonging to the population generated by the median-fit line \citep[see][]{Savino2022,Slaughter2025} is indicated by color. The calculated distance modulus, as well as the PWZ calibration coefficients for the JWST F090W and F150W bands are presented in the upper-left corner for reference. $\alpha_{JWST}$ is the combined standard offset and metallicity factor, and $b_{JWST}$ is the period coefficient.}
    \label{fig:jwstoldcalibration}
\end{figure}

Using the WLM-based calibration for the F090W and F150W filters from \cite{Slaughter2025}, we fit the PWZ to JWST data and find a distance modulus to Leo~P of $\mu = 26.01\pm0.04$ ($1.60\pm0.03$ Mpc), which is in agreement with the results reported in \autoref{sec:HSTpwz}, as well as the TRGB and RR Lyrae distances determined by \cite{McQuinn2015}. The PWZ fit can be seen in \autoref{fig:jwstoldcalibration}, where every source is well within reasonable expected uncertainty of the median fit line.

\subsection{Updated PWZ Calibration}

Using the HST-derived distance from \autoref{sec:HSTpwz} (in order to avoid circularity), we conduct a simultaneous calibration of the PWZ to the JWST F090W and F150W filters for both the present Leo~P sources and the 35 RRab sources with robust periods found in WLM by \citet{Slaughter2025}. We find PWZ calibration values of
\begin{equation*}
    \begin{split}
        \alpha &= -0.97^{+0.10}_{-0.11}\\
        b_{JWST}&= -0.79^{+0.48}_{-0.50}
    \end{split}
\end{equation*}

\begin{figure}
    \centering
    \includegraphics[width=\linewidth]{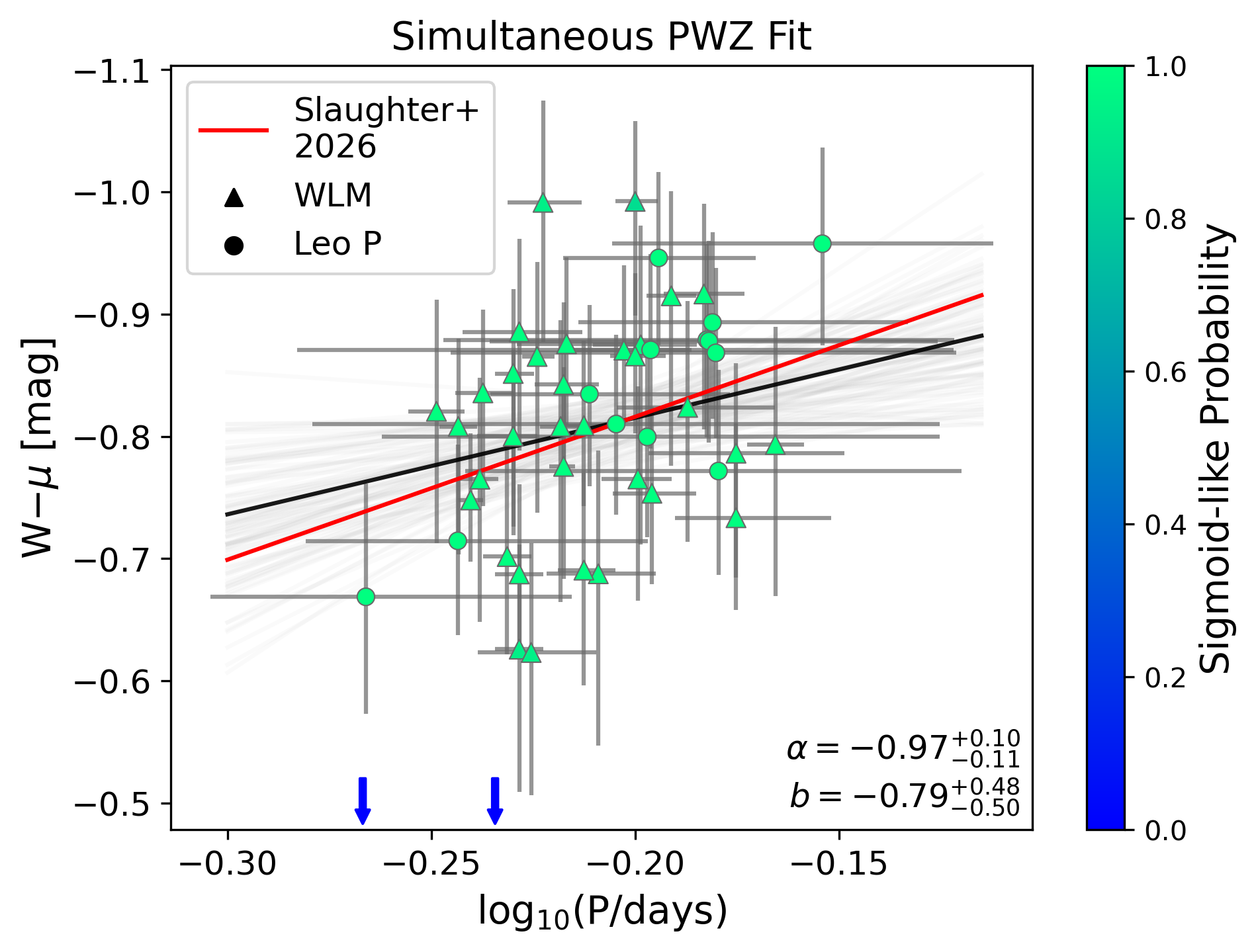}
    \caption{The PWZ calibration in the JWST filters using the Leo~P RRab data (circles), as well as the RRab WLM data (triangles) from \citet{Slaughter2025}. The fit line to both data sets is presented in black, while a random subset of other MCMC fits are in gray. The sigmoid-like probability of belonging to the population generated by the fit \citep[see][]{Savino2022,Slaughter2025} is indicated by color. For reference, the PWZ calibration to \textit{only} the WLM data, presented in \citet{Slaughter2025}, is shown in red. Two of the WLM sources fall outside the y-axis range, which has been reduced to better illustrate the difference between the WLM-only and combined fits. These sources are represented with arrows at their period locations, colored according to the same sigmoid-like probability. They both have very small calculated probabilities ($\sim 0.001$), and therefore do not have substantial impact on the PWZ fits.}
    \label{fig:jwstpwz_both}
\end{figure}

The final fit to both populations can be found in \autoref{fig:jwstpwz_both}. For ease of visualization, the y-axis for both data sets represents the difference between the Wesenheit magnitude (W) for each individual source minus the relevant host galaxy distance modulus ($\mu$). Again, the dark black line represents the new fit, and point color indicates the sigmoid-like probability of the source belonging to the population generated by the fit. The WLM-only fit from \citet{Slaughter2025} is shown in red, for reference. In order to better show the difference in the WLM-only and combined fits, the y-axis range has been reduced, cropping out two WLM sources from the figure. The x-axis locations of these points are shown with arrows, colored according to the same scheme as the rest of the data. The sigmoid-like probabilities of these sources belonging to the main population are determined to be very low ($\sim 0.001$), so they have little impact on the PWZ fits. This figure shows the somewhat greater range of periods identified in the Leo~P sources, helping to better anchor the slope of the fit than with the WLM data alone. We find the constraint on both $\alpha$ and $b_{JWST}$ improves when fitting to both data sets simultaneously, as compared to either in isolation.

In general, the uncertainty on the slope,  $b_{JWST}$, is the greater source of overall uncertainty on the fit. When comparing the WLM-only fit of \citet{Slaughter2025} to the combined PWZ calibration, we find that the fractional uncertainty (calculated with respect to the geometric mean of the asymmetric uncertainty values) on the slope ($b_{JWST}$) decreases from 94.0\% to 62.6\%. Similarly, the fractional uncertainty on the intercept ($\alpha$) decreases from 22.3\% to 10.8\%. 

\subsection{Metallicity Treatment}
It is important to discuss the treatment of RR Lyrae metallicity when fitting the PWZ to extragalactic stellar populations such as these. As it stands, the JWST distance calibration does not have an explicit metallicity dependence, where the metallicity term is wrapped into the intercept, as in \cite{Slaughter2025}:

\begin{equation}
    \alpha = a_{JWST} + c_{JWST}\textrm{[Fe/H]}.
\end{equation}

Meaning we are implicitly assuming that the RR Lyrae populations in Leo~P and WLM are comparable. In both \cite{Slaughter2025} and here \citep[as well as the HST-only work by][]{Savino2020}, the HST PWZ distance determinations to which the JWST calibration is anchored make use of a top-hat [Fe/H] prior over a physically-informed interval. This interval is chosen to be broad enough to capture the effect of our ignorance in the metallicity by way of the final distance (and, by extension, calibration) uncertainties. 

In general, external galaxies lack robust RR Lyrae metallicity measurements. Some assumption about the metallicity is therefore required. \cite{Slaughter2025} test a number of alternative treatments for the metallicity, including relationships with Fourier decomposition terms \citep[as in][]{Simon1982, Simon1988,Clement1992, Kovacs1995, Jurcsik1996, Morgan2003, Morgan2007, Kovacs2023, Li2023, Muraveva2025}, and ab-type periods \citep[as in][]{Sarajedini2023}. They find that none of the tested methods, when reasonable assumptions are employed, meaningfully change the final values of the PWZ calibration, nor do they see meaningful reduction of the uncertainties. As such, we conclude the assumed similarity between the populations is no more arbitrary than the other available methods.

\section{Discussion}\label{sec:discussion}

\subsection{HST and JWST Co-observation}\label{sec:co-obs}

We can use the RR Lyrae which are found in both the HST and JWST observations to better assess our methodology. We find 14 such co-observed sources. We compare the output features of the MCMC template-fitting for these 14 sources. Due to differences in filter set, the primary comparable values are the source period and pulsation type. Of the 14 sources observed in both HST and JWST, 12 are determined to be RRab, 2 RRc, with no disagreement. 

\begin{figure}
    \centering
    \includegraphics[width=\linewidth]{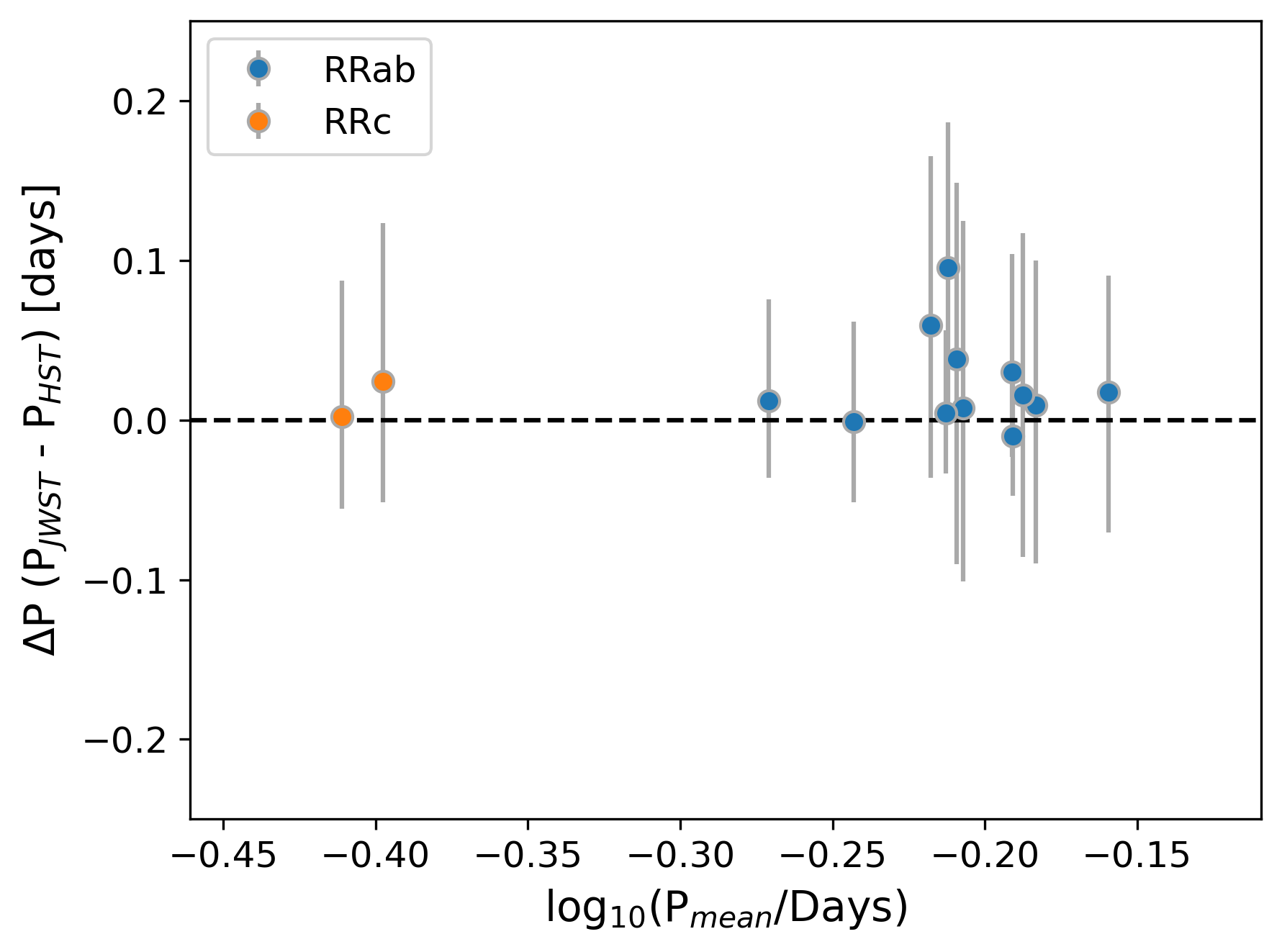}
    \caption{The difference in the MCMC best-fit periods derived from the JWST and HST data, for the co-observed sources. The expected value of the difference (0) is indicated by the dashed black line. The source type (which is agreed upon by the two fits in all cases) is indicated by color (RRab, blue; RRc, orange).}
    \label{fig:Pdiffs}
\end{figure}

To compare the periods, \autoref{fig:Pdiffs} shows the average of the two period estimates versus their difference ($\Delta P = P_{JWST} - P_{HST}$) for the 14 overlap sources. The uncertainties (added in quadrature) are shown, and the expected value (0) is indicated by a dashed horizontal line. Note that the presented uncertainties represent the 16th and 84th percentiles on the period posteriors from the MCMC. In this case, the posteriors on the JWST sampling exhibit heavier tails than a typical Gaussian distribution, increasing the uncertainties. This is a result of the shorter time-baseline of the JWST observations yielding only one cycle of data. Of the 14 sources, only one is not within the 1-$\sigma$ uncertainty of equivalent estimates.

\subsection{Oosterhoff Classification}
Generally, the mean period of the ab-type pulsators in an RR Lyrae population can be used as a first-pass method to assess the Oosterhoff classification of the system \citep[$\langle P_{ab}\rangle \approx 0.55d$ for Oosterhof I systems and $\langle P_{ab}\rangle \approx 0.65d$ for Oosterhof II systems,][]{Monelli2022}. However, the relatively large uncertainties in the JWST-derived periods described above warrant further analysis.

To start, for the ab-type pulsators, we find an average period in the HST data of $\langle P_{ab}\rangle = 0.613\pm0.037$, and in the JWST data of $\langle P_{ab}\rangle = 0.679\pm 0.067$, where the reported uncertainties simply represent the standard deviation of the mean. Alone, the JWST ab-type period mean indicates the RR Lyrae population of Leo~P may closely resemble that of an Oosterhoff II cluster. In contrast, the HST ab-type period mean places Leo P in the so-called ``Oosterhoff gap" between the Oosterhoff I (OoI) and II (OoII) classifications. This is common among extragalactic populations, in comparison to Galactic globular clusters\citep{Catelan2009, Monelli2022}. For example, based on its ab-type mean period, WLM also falls into the Oosterhoff gap \citep[$\langle P_{ab}\rangle = 0.61\pm0.03$;][]{Slaughter2025}.

Beyond ab-type period means, the Oosterhoff classification of an RR Lyrae population can be assessed with morphological analysis of the Bailey diagrams and CMD. \autoref{fig:hstbailey} and \autoref{fig:jwstbailey} show the empirical V and I band loci for the Oosterhoff classifications on the Bailey diagrams \citep[OoI, solid; OoII, dashed,][]{Kunder2013a, Kunder2013b, ArellanoFerro2015, Yepez2020}, where the V band is sufficiently similar to F475W, and the I band to F814W and F090W. In these, the RRab-type sources appear to fall largely between the two loci, with a possible skew toward the OoI locus (especially in the F475W data). In contrast, the observed RRc sources fall much more closely to the OoII loci, for both the linear (black) and parabolic (gray) solutions.

Further evidence can be found in the CMD HB morphology. Previous studies of Galactic globular clusters have found that \citep[with few exceptions,][]{ArellanoFerro2024b} OoII systems exhibit a clear separation between RRab and RRc sources on the HB at the first overtone red edge (FORE) of the instability strip \citep[e.g.,][]{ArellanoFerro2015, Yepez2020, ArellanoFerro2024a}. In contrast, the HST and JWST CMDs presented in \autoref{fig:hstcmd} and \autoref{fig:jwstcmd}, respectively, show no FORE segregation. Instead the RRab and RRc sources share the inter-order region of the instability strip. This is indicative of an OoI population.

\citet{Caputo1978} suggest that the two distinct morphologies seen in the HB may be due to differences in the CMD location of the starting point at the zero-age horizontal branch (ZAHB), impacting the original mode in which a given RR Lyrae pulsates. They further elaborate how the location of the ZAHB is dependent on the mass and chemical composition of a given star, and present this as an explanation for the two Oosterhoff classifications.

All combined, the ab-type period means, Bailey diagram morphologies, and CMD morphologies fail to unanimously support Leo P's membership in either of the Oosterhoff classifications. We are therefore led to conclude that the galaxy, like many of its extragalactic peers, likely falls in the Oosterhoff gap.

\section{Conclusions}\label{sec:conclsions}

Here, we present an analysis of the RR Lyrae pulsating variable star population in the low-metallicity dwarf galaxy Leo~P with both HST and JWST observations. Using existing variability identification and template-fitting classification methods, we identify a population of 20 RR Lyrae in the HST observations, and 15 RR Lyrae in the JWST observations. We conduct an independent fit of the \textit{Gaia}-consistent PWZ to the HST population in order to obtain a measurement of the distance. This estimate is then used as an input to calibrate the PWZ in the associated JWST filters. For this, we simultaneously fit to the Leo~P population and the RR Lyrae in WLM from \citet{Slaughter2025}. 

We recover the previously-identified population of Leo~P RR Lyrae in the HST data from \citet{McQuinn2015}. We find that the period estimates are within the margin of uncertainty from those previously reported for all but one source. Additionally, the estimate of the distance to Leo~P is in agreement to the value adopted by the previous study.

We build upon the original WLM-based JWST PWZ calibration, and find values within agreement, and with reduced error. We find that the slope of the fit is the greater source of uncertainty, compared to the intercept adjustment.

When comparing the MCMC-fit properties of the 14 co-observed sources, we find agreement between the HST and JWST data for their classifications, and consistency for their periods of pulsation.

Here, we show that the existing methodologies for identifying and classifying RR Lyrae are applicable to JWST IR data, provided the observations have sufficient phase-space coverage. While this poses potential difficulties for future JWST archival work (for which, the observation duration, and therefore phase coverage, is pre-determined), we highlight the continued success of the RR Lyrae identification scheme in both short- and long-duration data. This work demonstrates the near-full potential of extragalactic RR Lyrae period-fitting under standard JWST technical specifications. While period determination is essential for PWZ distancing, simple identification should be sufficient to probe ancient star formation. 


While some nearby galaxies will additionally be suited for other methods of standard candle distancing, such as Tip of the Red Giant Branch (TRGB) observations \citep[e.g.,][]{Conn2011, Conn2012, Lee2021}, many low-mass sources of interest will not have well-populated late-stage regions on the CMD, making TRGB methods unreliable. Here, we show that JWST data can be used to determine RR Lyrae-based distances, provided the observations are designed with the sufficient phase coverage. \citet{Savino2022} finds robust agreement between PWZ and TRGB distancing for galaxies with $M_V<-9.5$. This cutoff corresponds to an expected population of $\sim103$ RR Lyrae \citep[according to the empirical relation presented in][]{MartinezVazquez2019, MartinezVazquez2023}, plenty sufficient for distancing purposes. In short, for many of the archival observations for which TRGB fitting is untenable, we can still obtain distancing through RR Lyrae PWZ methods.

\section*{Acknowledgments}
This work is based on observations with the NASA/ESA/CSA James Webb Space Telescope obtained (program JWST-GO-1617) from the Data Archive at the Space Telescope Science Institute, which is operated by the Association of Universities for Research in Astronomy, Incorporated, under NASA contract NAS5-03127.

This work is also based on observations with the NASA/ESA Hubble Space Telescope (program GO-13376) obtained from the Data Archive at the Space Telescope Science Institute, which is operated by the Association of Universities for Research in Astronomy, Incorporated, under NASA contract NAS5-26555.

This research has made use of NASA's Astrophysics Data System Bibliographic Services. All of the data presented in this article were obtained from the Mikulski Archive for Space Telescopes (MAST) at the Space Telescope Science Institute. The specific observations analyzed can be accessed via \dataset[doi:10.17909/nhf1-6g39]{https://doi.org/10.17909/nhf1-6g39}.

\facilities{HST (ACS), JWST (NIRCam)}
\software{This research made use of routines and modules from the following software packages: 
\texttt{AstroPy} \citep{Astropy:2013,Astropy:2018,Astropy:2022},
\texttt{DOLPHOT} \citep{Dolphin2000, Dolphin2001, Dolphin2004, Dolphin2016, Weisz2024},
\texttt{emcee} \citep{Foreman-Mackey2013},
\texttt{Matplotlib} \citep{Matplotlib:2007},
\texttt{NumPy} \citep{Numpy:2020},
\texttt{Pandas} \citep{McKinney_2010, McKinney_2011},
\texttt{Psearch} \citep{Saha2017},
and \texttt{SciPy} \citep{Scipy:2020}
}

\appendix
\section{Extended Data}\label{sec:datatable}
An abbreviated table showing every source identified in this study in both HST and JWST data--as well as their on-sky locations, pulsation properties, and pulsation type--can be found in \autoref{tab:extendedtable}.

\begin{splitdeluxetable*}{cccccccccccBccccccccBcccccccc}
\caption{Multi-Catalog RR Lyrae Pulsation Parameters} \label{tab:extendedtable}
\tablehead{\colhead{JWST ID} & \colhead{HST ID} & \colhead{RA} & \colhead{Dec} & \colhead{Type} & \colhead{JWST Period} & \colhead{JWST eP-} & \colhead{JWST eP+} & \colhead{HST Period} & \colhead{HST eP-} & \colhead{HST eP+} & \colhead{McQ. Period} & \colhead{McQ. eP} & \colhead{F090W Amp} & \colhead{F090W eAmp-} & \colhead{F090W eAmp+} & \colhead{F090W Mag} & \colhead{F090W eMag-} & \colhead{F090W eMag+} & \colhead{ } & \colhead{F475W Amp} & \colhead{F475W eAmp-} & \colhead{F475W eAmp+} & \colhead{F475W Mag} & \colhead{F475W eMag-} & \colhead{F475W eMag+} & \colhead{ } \\ 
\colhead{ } & \colhead{ } & \colhead{ } & \colhead{ } & \colhead{ } & \colhead{$\mathrm{d}$} & \colhead{$\mathrm{d}$} & \colhead{$\mathrm{d}$} & \colhead{$\mathrm{d}$} & \colhead{$\mathrm{d}$} & \colhead{$\mathrm{d}$} & \colhead{$\mathrm{d}$} & \colhead{$\mathrm{d}$} & \colhead{$\mathrm{mag}$} & \colhead{$\mathrm{mag}$} & \colhead{$\mathrm{mag}$} & \colhead{$\mathrm{mag}$} & \colhead{$\mathrm{mag}$} & \colhead{$\mathrm{mag}$} & \colhead{ } & \colhead{$\mathrm{mag}$} & \colhead{$\mathrm{mag}$} & \colhead{$\mathrm{mag}$} & \colhead{$\mathrm{mag}$} & \colhead{$\mathrm{mag}$} & \colhead{$\mathrm{mag}$} & \colhead{ }}
\startdata
1041 & 3691 & 155.435 & -18.093 & ab & 0.70138 & 0.08333 & 0.06782 & 0.68411 & 0.0283 & 0.028 & 0.69 & 0.11 & 0.383 & 0.1119 & 0.1264 & 26.124 & 0.0416 & 0.0414 & ... & 0.505 & 0.2054 & 0.2525 & 26.835 & 0.0717 & 0.0724 & ... \\
1056 & 3848 & 155.442 & -18.084 & ab & 0.65917 & 0.04994 & 0.0728 & 0.62894 & 0.01333 & 0.0185 &  &  & 0.47 & 0.1319 & 0.11 & 26.151 & 0.0445 & 0.0451 & ... & 0.652 & 0.2427 & 0.2312 & 26.89 & 0.0818 & 0.0813 & ... \\
1060 & 3821 & 155.435 & -18.093 & ab & 0.6571 & 0.09785 & 0.09185 & 0.64151 & 0.04351 & 0.02591 &  &  & 0.324 & 0.0872 & 0.1217 & 26.142 & 0.0421 & 0.0419 & ... & 0.543 & 0.1658 & 0.2088 & 26.849 & 0.075 & 0.0728 & ... \\
1062 &  & 155.439 & -18.09 & ab & 0.65783 & 0.08148 & 0.08515 &  &  &  &  &  & 0.474 & 0.1268 & 0.1072 & 26.143 & 0.0397 & 0.0426 & ... &  &  &  &  &  &  & ... \\
1072 & 3863 & 155.441 & -18.093 & ab & 0.63655 & 0.12713 & 0.10896 & 0.59854 & 0.02002 & 0.01865 &  &  & 0.327 & 0.0896 & 0.1317 & 26.152 & 0.0401 & 0.0418 & ... & 0.84 & 0.2482 & 0.2108 & 26.843 & 0.0869 & 0.0851 & ... \\
1116 & 3726 & 155.436 & -18.094 & ab & 0.57084 & 0.04899 & 0.0613 & 0.57171 & 0.01346 & 0.01225 & 0.57 & 0.13 & 0.396 & 0.1154 & 0.1245 & 26.187 & 0.0432 & 0.0412 & ...  & 0.927 & 0.1855 & 0.1638 & 26.851 & 0.0739 & 0.0713 & ... \\
1117 & 3939 & 155.435 & -18.09 & ab & 0.63533 & 0.09542 & 0.10501 & 0.5759 & 0.01306 & 0.0089 & 0.61 & 0.06 & 0.516 & 0.1189 & 0.0889 & 26.184 & 0.0432 & 0.0457 & ... & 0.944 & 0.2297 & 0.1701 & 26.833 & 0.0718 & 0.0721 & ... \\
1120 & 3840 & 155.439 & -18.096 & ab & 0.6614 & 0.09457 & 0.09078 & 0.56613 & 0.01069 & 0.01045 & 0.57 & 0.04 & 0.5 & 0.1298 & 0.0962 & 26.203 & 0.0455 & 0.0473 & ... & 0.848 & 0.2308 & 0.1927 & 26.87 & 0.071 & 0.0711 & ... \\
... & ... & ... & ... & ... & ... & ... & ... & ... & ... & ... & ... & ... & ... & ... & ... & ... & ... & ... & ... & ... & ... & ... & ... & ... & ... & ... \\
\enddata

\tablecomments{An abbreviated table of the RR Lyrae pulsation parameters and types obtained for the 21 RR Lyrae sources in this study. Blank spaces indicate a given source was not in the associated set of observations. Included as well are the RA and Dec of each source, and periods obtained in \citet{McQuinn2015}, where applicable. These data are also available in
full in machine-readable format.}
\end{splitdeluxetable*}

\section{Full Phase-Folded Light Curves}\label{sec:fullLCs}
\begin{figure}
    \centering
    \includegraphics[width=\linewidth]{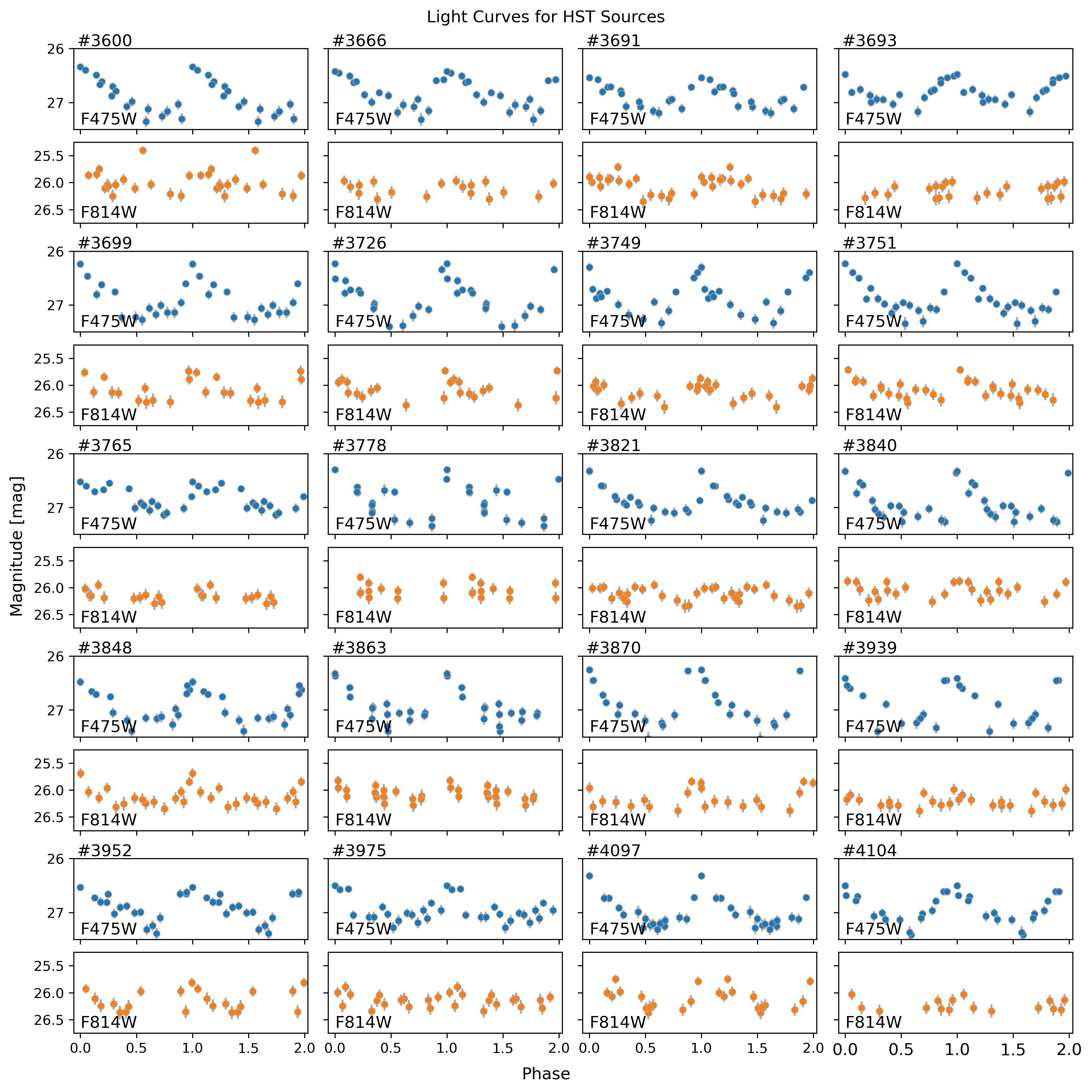}
    \caption{The full set of phase-folded light curves from the RR Lyrae identified in the HST data herein. Both the F475W (blue) and F814W (orange) data are shown for each source. The magnitude axes have a range of 1.5 mag for both filters, in order to best show the variability while emphasizing the differences in amplitude between the filters without obscuring any epochs. Two cycles are shown for viewing convenience.}
    \label{fig:HSTlcFull}
\end{figure}

\begin{figure}
    \centering
    \includegraphics[width=\linewidth]{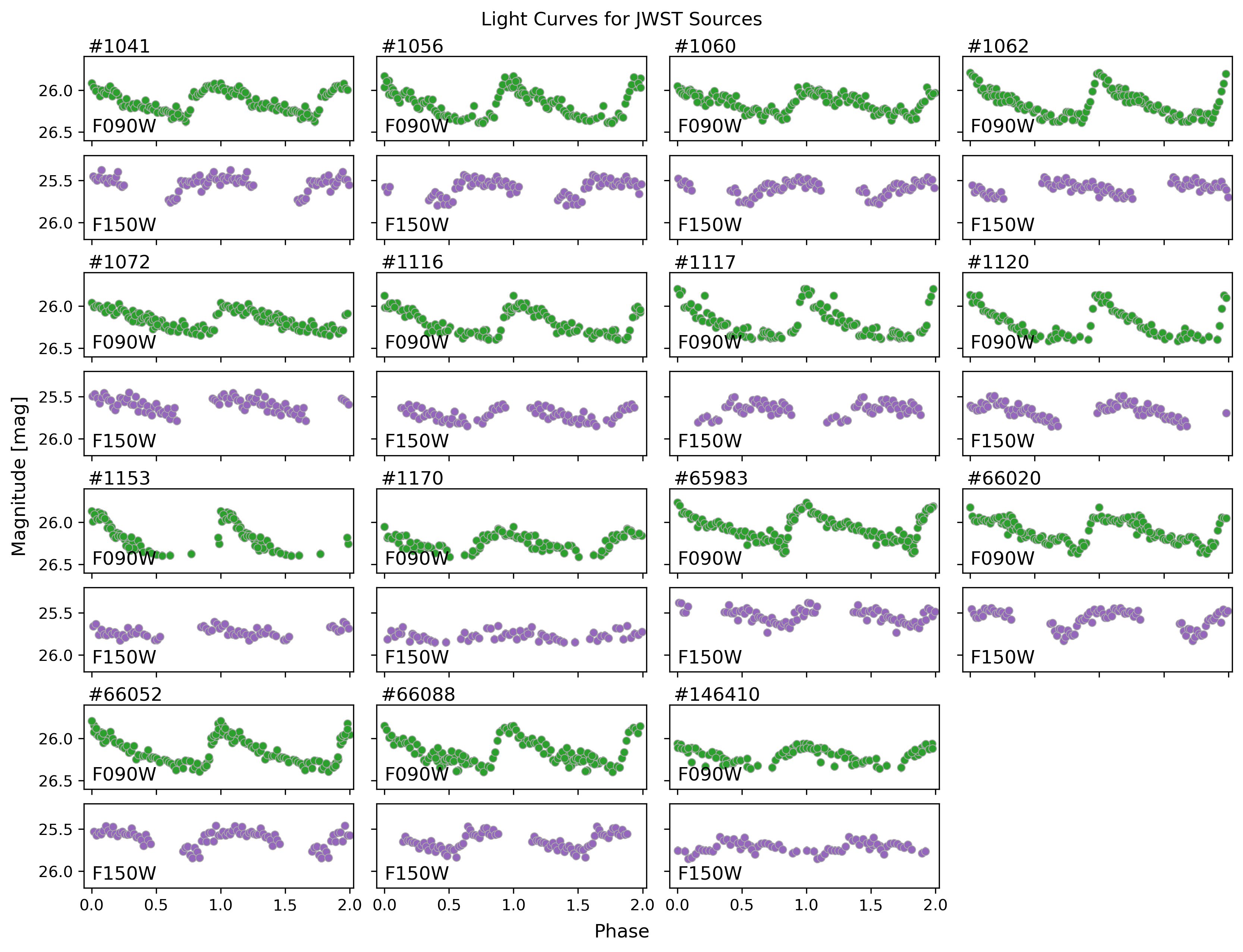}
    \caption{The full set of phase-folded light curves from the RR Lyrae identified in the JWST data herein. Both the F090W (green) and F150W (purple) data are shown for each source. The magnitude axes have a range of 1 mag for both filters, in order to best show the variability while emphasizing the differences in amplitude between the filters without obscuring any epochs. Two cycles are shown for viewing convenience.}
    \label{fig:JWSTlcFull}
\end{figure}

The full sets of light curves for the HST- and JWST-identified sources can be found in \autoref{fig:HSTlcFull} and \autoref{fig:JWSTlcFull}, respectively.

\bibliography{full.bib}{}
\bibliographystyle{aasjournalv7}

\end{document}